\documentclass[twocolumn]{aastex631}
\usepackage{amsmath}
\usepackage{mathrsfs}

\graphicspath{{fig/}}
\defcitealias{Hou2024}{HY24}

\begin{document}
%\title{The Effects of Dust Diffusion on Streaming Torque}
\title{Streaming Torque with Turbulent Diffusion}

\correspondingauthor{Cong Yu}
\email{yucong@mail.sysu.edu.cn}

\author[0000-0002-8125-7320]{Qiang Hou}
\affiliation{School of Physics and Astronomy, Sun Yat-sen University, Zhuhai 519082, China}
\affiliation{CSST Science Center for the Guangdong-Hong Kong-Macau Greater Bay Area, Zhuhai 519082, China}
\affiliation{State Key Laboratory of Lunar and Planetary Sciences, Macau University of Science and Technology, Macau, China}
\author[0000-0003-0454-7890]{Cong Yu}
\affiliation{School of Physics and Astronomy, Sun Yat-sen University, Zhuhai 519082, China}
\affiliation{CSST Science Center for the Guangdong-Hong Kong-Macau Greater Bay Area, Zhuhai 519082, China}
\affiliation{State Key Laboratory of Lunar and Planetary Sciences, Macau University of Science and Technology, Macau, China}

\begin{abstract}
Fast type-I migration of (proto)planets poses a challenging problem for the core accretion formation scenario. We found that the dust-induced ``Streaming Torque (ST)'' may slow down or even reverse the planet migration in \cite{Hou2024}. But in realistic protoplanetary disks, dust diffusion induced by gas turbulence may have important influences on ST. We perform linear analysis to investigate the effects of dust diffusion on ST. The dependence of ST on the dust diffusion may provide better constraints on the turbulence strength and the stopping time $\tau$. We derive the dispersion relation for all the wave modes in the two-fluid system. The dust diffusion will smooth the short-wavelength structure of the the quasi-drift mode and split it into two predominant D-drift modes with opposite directions.  The outgoing D-drift mode will contribute to a negative torque on planets, particularly when $\tau \sim 0.1$, which slightly shifts the zero-torque turning point. We explore how ST depends on the regimes of aerodynamic drag, dust mass fraction and disk scale height. We compare the radial wavenumbers of D-drift modes under different formulations of dust diffusion and find qualitative agreement. In all cases, $\tau$ at the zero-torque turning point, which determines the direction of planetary migration, consistently remains on the order of $\sim 0.1$, corresponding to large pebble-sized dust grains. This suggests that rapid dust coagulation can inhibit the inward migration of planets, implying that weak gas turbulence may enhance the survival of protoplanets.
\end{abstract}


\section{Introduction} \label{introduction}

Recently, we found a new streaming torque (ST) \citep{Hou2024} (\citetalias{Hou2024} hereafter) to address the problem of inward and rapid Type-I migration \cite[]{Goldreich1979,Korycansky1993,Ward1997,Tanaka2002,D'Angelo2010}. ST arises from the quasi-drift mode, which is driven by the inward drift of dust grains. In \citetalias{Hou2024}, we employed linear analysis based on the Nakagawa–Sekiya–Hayashi (NSH) equilibrium, while ST is universal in dust-gas coupled protoplanetary disks (PPDs) beyond that. In such systems, the dust tends to follow Keplerian motion in the azimuthal direction, while the gas is sub-Keplerian due to the pressure gradient. This results in aerodynamic drag between the dust and the gas, causing the dust to be pulled by the gas, losing angular momentum (AM) and drifting inward. Along the dust streamline, planets can induce the quasi-drift mode, resulting in ST which is positive in most cases. Our calculations show that the dust plays a significant role in planetary migration, even when it occupies only a small mass fraction of PPDs. Both the dusty wake in an asymmetric pattern leading the planet and disk torques are consistent with previous hydrodynamic simulations \citep{Llambay2018,Regaly2020,Chrenko2024}, which confirms the existence of ST further.

Because of the small drift velocity of dust, the length-scale of the quasi-drift mode is also very small, similar to the modes inducing streaming instability (SI) \citep{Youdin2005,Magnan2024a}. The drift velocity of dust can resonate with other motions, such as with the epicyclic motion leading to SI. This instability family is called ``Resonant Drag Instability (RDI)'' \citep{Squire2018a,Squire2018b}. If the drift velocity is large enough, the quasi-drift mode can also resonate with density waves (DWs) \citep{Hopkins2018,Magnan2024b}, which is unrealistic in PPDs.

However, gas turbulence is ubiquitous in PPDs and can be induced by various hydrodynamic instabilities. Previous studies find that it can influence the horseshoe drag \cite[e.g.][]{Baruteau2010} but not on Lindblad torque (LT), equivalent to the unsaturated effects on corotation torque (CT) by viscosity. Other hydrodynamic simulations also found viscosity could slower the inward migration of planets \cite[e.g.][]{Yu2010}. The determination of turbulence intensity has always been of great importance, not only for the planetary migration problem we are concerned with here.

Dust diffusion is driven by gas turbulence, which tends to smooth out small-scale structures, as demonstrated in several studies on the SI in turbulent gas \citep[e.g.,][]{Chen2020, Gole2020, Umurhan2020}. These studies have shown that dust diffusion suppresses the growth rate of SI, potentially rendering it ineffective in certain PPD environments. Consequently, alternative mechanisms for dust growth, such as coagulation instability \citep{Tominaga2021,Tominaga2022a,Tominaga2022b} and dusty Rossby wave instability \citep{Liu2023}, have been proposed. However, its effects on dusty torque have not been concluded \citep{Llambay2018,Regaly2020,Chrenko2024}. The quasi-drift mode, which generates ST, also exhibits a short-wavelength structure that is susceptible to dust diffusion. In this paper, we aim to investigate the impact of dust diffusion on ST, which is highly sensitive to the stopping time between dust and gas. The final results for planetary migration can then provide constraints on the turbulent strength in turn, which is significant in the context of core accretion studies \citep{Lee2024}.

The paper is organized as follows. In Section \ref{method}, we introduce our methodology, including the fundamental equations of gas and dust motion, linear perturbation theory, and the numerical method, which is largely the same as in \citetalias{Hou2024}. In Section \ref{wave}, we analyze the modes in this two-fluid system, modified by dust diffusion, using both analytical and numerical approaches. Section \ref{torque} presents the results of the disk torque on planets across a broad parameter space. We provide further discussion in Section \ref{discussion}, followed by the summary and conclusions in Section \ref{conclusion}.
 
\section{Methodology} \label{method}
In this section we will introduce our framework on a two-dimensional (2D) dust-gas coupled PPD based on \citetalias{Hou2024}, where a protoplanet is embedded. The planet acts as a perturbation source in the disk. We then perform a linear analysis, and due to the complexity of the resulting equations, they are solved numerically.

\subsection{Basic Setup of PPD with Planet}
In the dust-gas coupled PPD, we assume an embedded planet with low mass $M_p$ is in a circular orbit with distance $r_p$, centered by a single host star with mass $M_{\star}$. And it is moving with Keplerian frequency $\Omega_p = \Omega_{\rm{K}}(r_p)$. Specifically, $M_p$ is assumed to be below a few percent of the thermal mass, $M_{\rm{th}} = h_p^3 M_{\star}$ \citep{Goodman2001}, beyond which the planet would significantly affect its surroundings and open gaps in the disk. Here, $h_p$ is the aspect ratio of the PPD at $r_p$, defined as $h_p \equiv H_p/r_p$, where $H_p$ is the gas scale height. Typically, the value of $h_p$ ranges from $0.03$ to $0.10$, and we adopt $h_p= 0.03$ as the fiducial value for our calculations. The planet is fixed on a circular orbit, 

We place the planet at the origin of a non-inertial Cartesian coordinate system, defined as $x = (r - r_p)/H_p$ and $y = r_p (\theta - \Omega_p t)/H_p$, while neglecting all curvature terms, i.e., using the ``shearing sheet approximation'' \citep{Goldreich1965}.

\subsection{Gas Equations}
The 2D governing equations for gaseous motion read
\begin{gather} 
    \frac{\partial \Sigma_g}{\partial t}+\nabla \cdot\left(\Sigma_g \boldsymbol{u}_g\right) =0,  \label{eq_gas_1} \\
    \frac{\partial \boldsymbol{u}_g}{\partial t}+\boldsymbol{u}_g \cdot \nabla \boldsymbol{u}_g= - 2 \boldsymbol{\Omega}_p \times \boldsymbol{u}_{g} - \nabla \left( \Phi_{p} + \Phi_{\rm{eff}} \right) \nonumber \\
    +\frac{\Sigma_d}{\Sigma_g} \frac{\boldsymbol{u}_d-\boldsymbol{u}_g}{t_{\text {s}}}-c_s^2 \nabla \ln \Sigma_g + 2\eta \Omega_p^2 r_p \boldsymbol{e}_{x}. \label{eq_gas_2}
\end{gather}
Here, $\Phi_p$ represents the planetary potential, and $\Phi_{\rm{eff}} = -S \Omega_p x^2$ is the effective (or tidal) potential, with $S$ as the orbital shear rate ($S = 1.5$ in our (sub)Keplerian disk model). Subscripts ``$g$'' and ``$d$'' indicate quantities for gas and dust, respectively. We adopt an isothermal equation of state. $c_s = \sqrt{P/\Sigma_g}$ is the isothermal sound speed, where $\Sigma_g$ is the gas  surface density and $P$ is the global surface pressure \citep{Paardekooper2020,Klahr2021}. The characteristic scale height of the gas is defined as $H_p \equiv c_s / \Omega_p$. The parameter $\eta = -\left( \partial P/ \partial \ln r \right)/\left(2\Sigma_g V_{\rm{k}}^2 \right) \simeq 0.5h_p^2 $ governs the sub-Keplerian nature of the disk. The third term on the right-hand side represents the aerodynamic drag between dust and gas, quantified by the stopping time $t_s$, which measures how long it takes for dust and gas to behave the same. The dimensionless quantity $\tau \equiv \Omega_p t_s$ is referred to as the Stokes number in turbulence-free PPDs. 

We assume the disk exhibits moderate gas turbulence, modeled using the $\alpha$-prescription \citep{Shakura1973}, with $\alpha$ ranging from $10^{-4}$ to $10^{-3}$. However, we neglect viscous forcing in the gaseous equations, like some previous works \citep[e.g.][]{Klahr2021,Tominaga2023}, as it is weak compared to the pressure in weakly or moderately turbulent PPDs. Since we primarily focus on dust dynamics, the only effect of turbulence is the induction of dust diffusion.

\subsection{Dust Diffusion Equations}
Dust diffusion induced by turbulent stirring can be modeled in several ways. Previous studies typically introduce an additional term, following Fick's law, into the dust continuity equation \citep[e.g.][]{Takahashi2014,Shadmehri2016,Latter2017,Dullemond2018,Chen2020,Umurhan2020}, while leaving the momentum equation unchanged. The governing equations for dust are given by:
\begin{gather}
    \frac{\partial \Sigma_d}{\partial t}+\nabla \cdot\left(\Sigma_d \boldsymbol{u}_d\right) = \nabla \cdot \left[ D\Sigma_g \nabla \left( \frac{\Sigma_d}{\Sigma_g} \right) \right], \label{eq_dust_in1} \\ 
    \frac{\partial \boldsymbol{u}_d}{\partial t}+\boldsymbol{u}_d \cdot \nabla \boldsymbol{u}_d= - 2 \boldsymbol{\Omega}_p \times \boldsymbol{u}_{d} - \nabla \left( \Phi_{p} + \Phi_{\rm{eff}} \right) - \frac{\boldsymbol{u}_d-\boldsymbol{u}_g}{t_{\text {s}}}. \label{eq_dust_in2}
\end{gather}
Here, $D=\alpha c_s H_p$ represents the diffusion coefficient induced by gas turbulence.

However, recent studies have shown that this formulation can lead to certain inconsistencies, such as the non-conservation of AM, as pointed out by \cite{Tominaga2019}. Consequently, new formulations have been proposed and implemented \citep{Tominaga2019,Klahr2021,HPH2022}, where both the continuity and momentum equations are revised. In this paper, we adopt the formulation of \cite{Klahr2021}, which decomposes the advected velocities into mean flow velocities and diffusion velocities:
\begin{gather}
    \boldsymbol{u}^{*} = \boldsymbol{u} + \boldsymbol{u}_{\rm{diff}}, \\
    \boldsymbol{u}_{\rm{diff}} = - D \frac{\Sigma_g}{\Sigma_d} \nabla \left( \frac{\Sigma_d}{\Sigma_g} \right).
\end{gather}
Then the dust equations read
\begin{gather} 
    \frac{\partial \Sigma_d}{\partial t}+\nabla \cdot\left(\Sigma_d \boldsymbol{u}_d^{*}\right) =0,  \label{eq_dust_1} \\
    \frac{\partial \boldsymbol{u}_d^{*}}{\partial t}+\boldsymbol{u}_d^{*} \cdot \nabla \boldsymbol{u}_d^{*}= - 2 \boldsymbol{\Omega}_p \times \boldsymbol{u}_{d}^{*} - \nabla \left( \Phi_{p} + \Phi_{\rm{eff}} \right) \nonumber \\
    -\frac{\boldsymbol{u}_d^{*}-\boldsymbol{u}_g}{t_{\text {s}}} - \frac{D \Sigma_g}{t_s \Sigma_d} \nabla \left( \frac{\Sigma_d}{\Sigma_g}\right). \label{eq_dust_2}
\end{gather}
Note that in Equation \eqref{eq_gas_2}, $\boldsymbol{u}_d$ is replaced by $\boldsymbol{u}_d^{*}$ under this formulation. In Equation \eqref{eq_dust_2}, the last term, diffusion term, on the right-hand side is added in a pressure-like form, distinguishing it from the traditional formulation. We choose to make this term proportional to the gradient of the dust-to-gas ratio, $\nabla \left(\Sigma_d/\Sigma_g\right)$, rather than the dust density gradient, $\nabla \Sigma_d$, as presented in \cite{Klahr2021}. This adjustment reflects that it is the dust concentration being diffused. However, the difference between these two gradients is negligible in our study. For further details and justification, see Appendix B of \cite{Klahr2021}.

Although Equations \eqref{eq_dust_in1}-\eqref{eq_dust_in2} exhibit inconsistencies, the underlying physical mechanism remains the same. In some cases, the differences between the two formulations are negligible \citep{Umurhan2020}. We will compare these two approaches, i.e., Equations \eqref{eq_dust_in1}-\eqref{eq_dust_in2} and Equations \eqref{eq_dust_1}-\eqref{eq_dust_2}, in Section \ref{model}.

\subsection{Linear Analysis} \label{linear}
Up to now, we have the complete set of two-fluid hydrodynamic equations, namely Equations \eqref{eq_gas_1}, \eqref{eq_gas_2}, \eqref{eq_dust_1}, and \eqref{eq_dust_2}. For readability, we will drop the superscript ``$*$'' from $\boldsymbol{u}_d^{*}$ in the subsequent discussion. In the absence of the planetary potential, an axisymmetric steady-state solution can be derived from these equations, commonly referred to as the NSH equilibrium \citep{Nakagawa1986},
\begin{gather}
    u_{dx,0} = - f_g \chi_1 \eta h_p^{-1} \Omega_{p}, \label{NSH_1} \\
    u_{dy,0} = - \frac{3}{2}\Omega_p x - f_g \chi_2 \eta h_p^{-1} \Omega_{p},  \label{NSH_2} \\
    u_{gx,0} = f_d \chi_1 \eta h_p^{-1} \Omega_{p},  \label{NSH_3} \\
    u_{gy,0} = - \frac{3}{2}\Omega_p x + \left( f_d \chi_2 - 1 \right) \eta h_p^{-1} \Omega_{p}. \label{NSH_4} 
\end{gather}
Here $f_{d/g} \equiv \Sigma_{d/g}/( \Sigma_{d} + \Sigma_{g})$ represents the mass fraction of dust/gas. Additionally, we define $\chi_1 \equiv 2f_{g}\tau/ [ 1+\left( f_{g} \tau \right)^2 ]$ and $\chi_2 \equiv 1/[1+ \left( f_{g} \tau \right)^2]$.

We apply linear analysis to examine Equations \eqref{eq_gas_1}, \eqref{eq_gas_2}, \eqref{eq_dust_1}, and \eqref{eq_dust_2} in a standard way \citep[e.g.][]{Narayan1987,Rafikov2012}, where all perturbation quantities are expressed as integrals in Fourier space, i.e., $ X_{1}(x,y) = \int_{-\infty}^{+\infty} \delta X(x,k_y) \exp \left( ik_y y \right) d k_y$ ($k_y$ is the azimuthal wavenumber). The final perturbation equations are presented in Appendix \ref{appd_Epstein}, where Equation \eqref{linear1}-\eqref{linear6} derived in the Stokes regime are solved numerically as our fiducial model. The equations derived in the Epstein regime are given by Equation \eqref{e_linear2}-\eqref{e_linear6}, while the difference on torque for the different regime is none (see Section \ref{tau}). Lastly, we use the following normalization
\begin{gather}
    \boldsymbol{w}_s = \Omega_p \tilde{\boldsymbol{w}}_s, \\
    \boldsymbol{u}_{d/g} = \Omega_p \tilde{\boldsymbol{u}}_{d/g}, \\
    \delta \boldsymbol{u}_{d/g} = q h_p^{-3} \Omega_p \delta \tilde{\boldsymbol{u}}_{d/g}, \\
    \delta \Sigma_{d/g} = q h_p^{-3} \Sigma_p \delta \tilde{\Sigma}_{d/g}.
\end{gather}
The parameters $\mu$ and $q$ represent the mass ratios of dust to gas $f_d/f_g$ and planet to star $M_{p}/M_{\star}$, respectively. $\boldsymbol{w}_s \equiv \boldsymbol{u}_d - \boldsymbol{u}_g = ( w_{s,x}, w_{s,y})$ represents the relative velocity between dust and gas, and will henceforth be referred to as the drift velocity.

\subsection{Numerical Methods}
We employ the numerical method outlined in \citetalias{Hou2024} to solve Equations \eqref{linear1}-\eqref{linear6} and obtain the perturbations induced by a low-mass planet. The core of this method is the relaxation technique \citep{Press1992,Huang2022}, which aims to solve the two-point boundary value problems through iterations. At the boundaries, we allow wave flow to exit freely by employing a WKB approximation process. To ensure sufficient convergence, we utilize a high resolution of $2\times10^{-4} H_p$, achieving an error less than $10^{-9}$. In practice, such a high resolution is not necessary for most of the parameter space when dust diffusion exists. The Smoothing length of planetary potential is set as $1/8H_p$ for a comparison with \citetalias{Hou2024}, and a detailed discussion about that is presented in Section \ref{soft}.

\section{Wave Propagation Properties} \label{wave}
In this section, we will focus on the wave propagation properties modified by dust diffusion in comparison to the results \citetalias{Hou2024}. Before delving into these results, we will perform an analysis of the dispersion relation.

\subsection{Dispersion Relation} \label{DR}
We present the eigenmodes excited in the dust-gas coupled system through analyzing its dispersion relations under three assumptions:
\begin{enumerate}
    \item The perturbations are nearly axisymmetric, which is suitable for the density waves far away from corotation resonance point and naturally fits the quasi-drift mode.
    \item The gas diffusion terms are ignored. Gas diffusion is not important because of the long-scale nature of gas waves.
    \item The gas is not damped by the dust because of the small dust-gas ratio $\mu$, while we reserve its driving terms on dust to relate their perturbations. 
\end{enumerate}
The detailed derivations and illustrations are given in Appendix \ref{appd_DR}.

We identify six important modes. The first two correspond to gaseous density waves, described by the  equation $1 - \tilde{\omega}_g^2 + k_x^2=0$ (normalized by $\Omega_p^2$). The next two are dusty density waves, which are presented by $ 1 - \tilde{\omega}_d^2 + f\left( \tau \right) = 0$. This relation closely resembles that of gaseous density waves, with the resilience term replaced by aerodynamic drag, as represented by $f(\tau)$.

The left two are related to the quasi-drift mode confirmed in \citetalias{Hou2024}. After some simplifications, the dispersion relation have the common term
\begin{gather}
    \tilde{\omega}_d^{\prime 3} - \left( \alpha k_x^2 + 1 \right) \tilde{\omega}_d^{\prime } - i \alpha k_x^2 = 0, \label{D-drift}
\end{gather}
{where three modes (solutions) can be obtained, two of which are predominant. In fact, these modes are split from the quasi-drift mode and have reduced wavenumbers. It is easy to see that without dust diffusion, Equation \eqref{D-drift} reproduces the free drift mode $\tilde{\omega}_d^{\prime}= \tilde{\omega}_d - k_x v_{dx} = 0 $ (the quasi-drift mode can then be recovered after adding gaseous damping). So, dust diffusion will split the original   quasi-drift mode into two predominant modes, which we we refer to as ``D-drift mode'' hereafter.  One of them is ingoing, similar to the quasi-drift mode, contributing a significant positive torque to the planet in most cases. In contrast, the other one is outgoing, resulting in a moderate negative torque on the planet. A detailed analysis will be provided in Section \ref{wavenumber}.
\subsection{Waves Excited by Planet}
\begin{figure*}[htbp]
    \centering
    \includegraphics[width=1.0\textwidth]{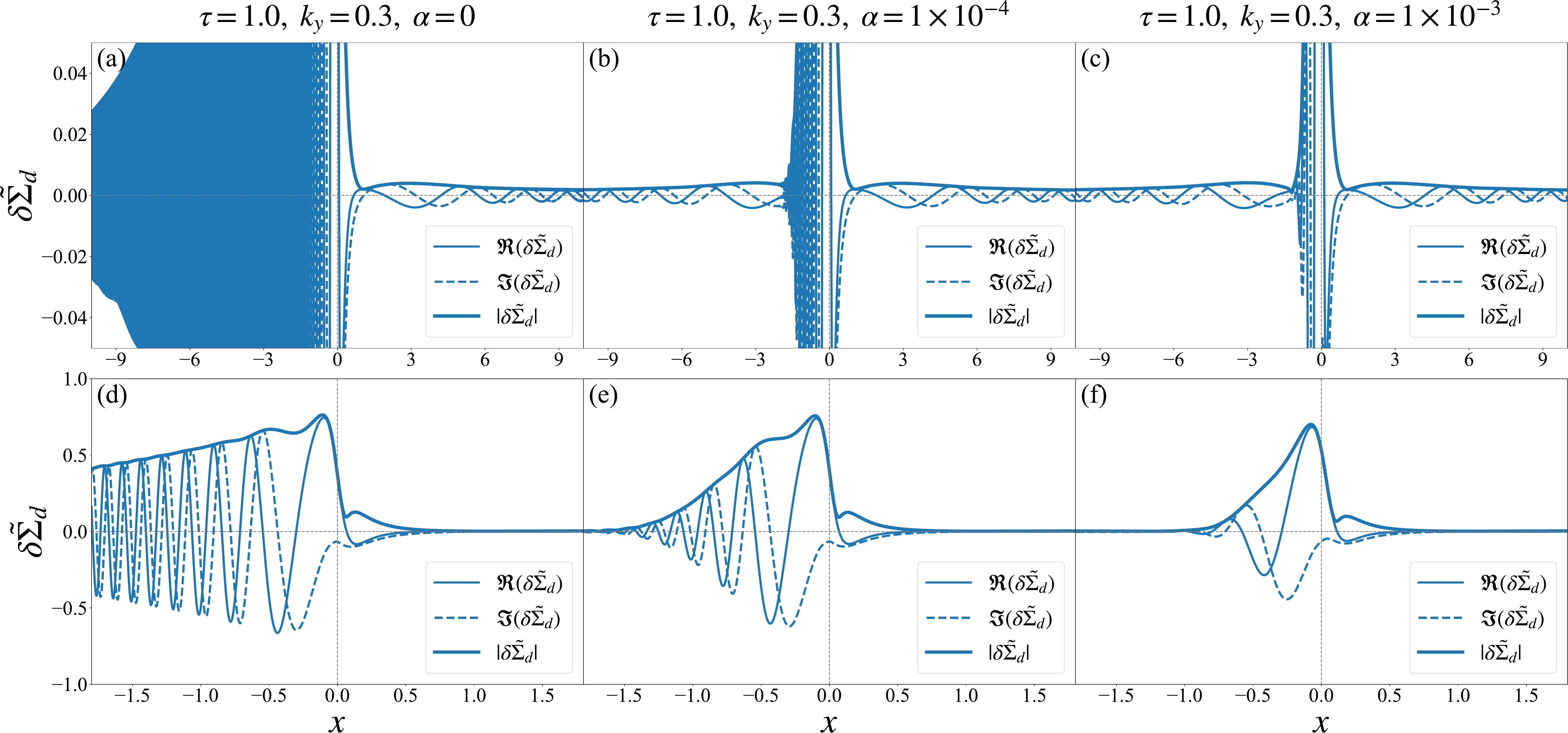}  
    \caption{Normalized dust density perturbations $\delta \tilde{\Sigma}_d$ with $f_d=0.01,\tau=1.0,k_y=0.3$. The panels in three columns display those with different $\alpha$ values. Form left to right, $\alpha=0$ (corresponding to Figure 1 in \citetalias{Hou2024}), $\alpha=1\times10^{-4}$ and $\alpha=1\times10^{-3}$. The panels in the second row display $\delta \tilde{\Sigma_d}$ same as those in the corresponding column, but with a larger $y$ range for a clear illustration of D-drift modes.}
    \label{fig_wave}
\end{figure*}
Because of a low $f_d$, gaseous waves are minimally influenced by dust, as confirmed in \citetalias{Hou2024}.

In Figure \ref{fig_wave} we display the wavy behaviors of the Fourier harmonics of the normalized dust density perturbation $\delta \tilde{\Sigma}_d $ as a function of $x$ for fixed $\tau=1.0,~ k_y=0.3$ and three different $\alpha$. Panels (a) and (d) with $\alpha=0$ correspond to Figure 1 in \citetalias{Hou2024}. Inside the planet ($x < 0$), the short length scale structure is caused by the quasi-drift mode, as a confirmed result of the appearance of dust drift velocity \citep{Hou2024,Hopkins2018}. In the second column, i.e., panel (b) and (e), it is evident that D-drift mode (corresponding to the original quasi-drift mode) has a larger length scale and experiences stronger damping as it propagates inward. With the same settings but a non-zero $\alpha$, panel (b) shows that the mode only extends to about $x=-2$ and nearly vanishes, becoming dominated by dusty density waves. This result aligns with our physical intuition, i.e., dust diffusion tends to smooth the density gradient. Panel (c) and (f) show the same trend, with a larger length scale, stronger damping, and a smaller wave amplitude.

\subsection{Wakes of Dust}
\begin{figure*}[htbp]
    \centering
    \includegraphics[width=1.0\textwidth]{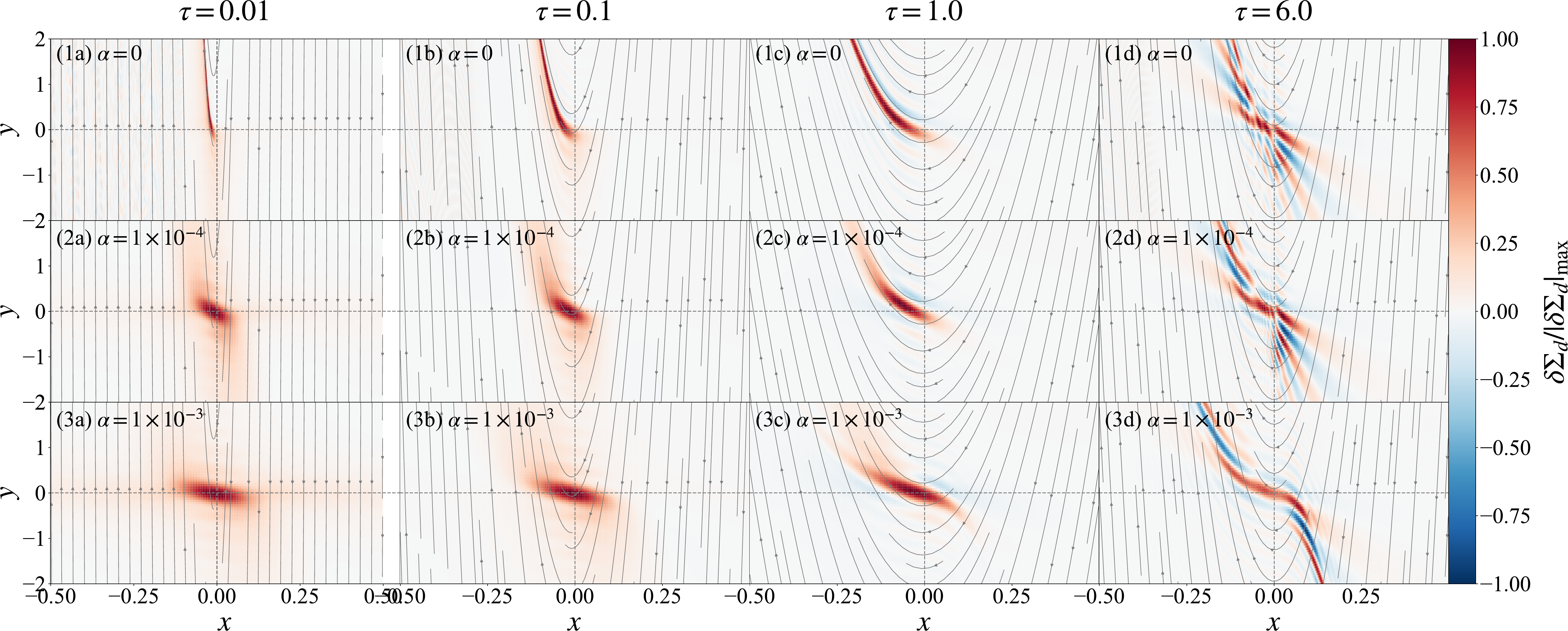}  
    \caption{Dusty wakes induced by the quasi-drift mode with $\alpha=0$ and D-drift modes with $\alpha \neq 0$. The first row with $\alpha = 0$ corresponds to the results in \citetalias{Hou2024}. The last two rows corresponds to the results with $\alpha = 1\times10^{-4}$ and $\alpha = 1\times10^{-3}$, respectively. From the leftmost to the rightmost column, $\tau = 0.01,0.1,1.0,6.0$, respectively. The solid gray lines with arrows in each panel represent the unperturbed dust streamlines given by Equation \eqref{NSH_1} and \eqref{NSH_2}.}
    \label{fig_wake}
\end{figure*}
We integrate the surface density perturbations in $k_y$ space and visualize them in real space. The gas in the disk exhibits the typical density waves discussed in \citetalias{Hou2024}, which we will not repeat here. Dust located far from the planet also shows density waves, while the distinct D-drift mode dominates in regions close to the planet. Therefore, we present only the dusty wake caused by the D-drift mode in this section.

The results are displayed in Figure \ref{fig_wake}, which compares the dusty wakes of planets for different values of $\tau$ and $\alpha$. The first row, with $\alpha = 0$, corresponds to the results in \citetalias{Hou2024}. The last two rows correspond to the cases with $\alpha = 1\times10^{-4}$ and $\alpha = 1\times10^{-3}$, respectively. From the leftmost to the rightmost column, $\tau = 0.01, 0.1, 1.0, 6.0$. The solid gray lines with arrows in each panel represent the unperturbed dust streamlines given by Equation \eqref{NSH_1} and \eqref{NSH_2}.

When $\alpha = 0$ and $\tau = 0.01, 0.1, 1.0$, the wake clearly aligns with the dust streamlines in each panel, illustrating the origin of the quasi-drift mode. Fixing $\tau$ and comparing the results with non-zero $\alpha$ values, we observe that the dusty wakes are smoothed by dust diffusion, similar to a convolution process. The length scale of the D-drift modes is larger than that of the quasi-drift mode, especially when $\tau$ is small. The wakes no longer follow the dust streamlines. Furthermore, the effect becomes more pronounced as $\alpha$ increases. 

In \citetalias{Hou2024}, we demonstrated that the quasi-drift mode is ingoing and predominantly appears in the second quadrant, as shown in panels (1a)-(1c). However, when $\alpha \neq 0$, the dusty wake becomes equally clear in the fourth quadrant. The wake outside the planet is primarily caused by another D-drift mode, which propagates in the opposite direction (outgoing), as discussed in Section \ref{DR}. Panels (2a)-(2c) and (3a)-(3c) demonstrate that the outgoing D-drift mode is prominent when $\tau$ is low, highlighting its significance in this regime. It is also easy to infer that the outgoing D-drift mode contributes a negative torque, in contrast to the ingoing mode. We will further demonstrate this in Section \ref{torque}.

When $\tau = 6.0$, multiple wakes appear on both the leading and trailing sides. In this regime, weaker gas damping results in more distinct stripes compared to those observed at $\tau = 1.0$. In panel (1d), the wake along the streamline intersects with others, causing amplitude oscillations near the planet, which are also evident in panel (2d). These oscillations can be smoothed out by dust diffusion, as shown in panel (3d). Interestingly, trailing wakes are visible in the fourth quadrant even when $\alpha = 0$. All these wakes are attributed to D-drift modes. However, their exact origins remain unclear. We guess those wakes can evole to dust deficits \citep{Llambay2018} through nonlinear effects. We plan to investigate these phenomena through further calculations, which may help clarify and verify these features.

\subsection{Radial Wavenumber} \label{wavenumber}
\begin{figure*}[htbp]
    \centering
    \includegraphics[width=1.0\textwidth]{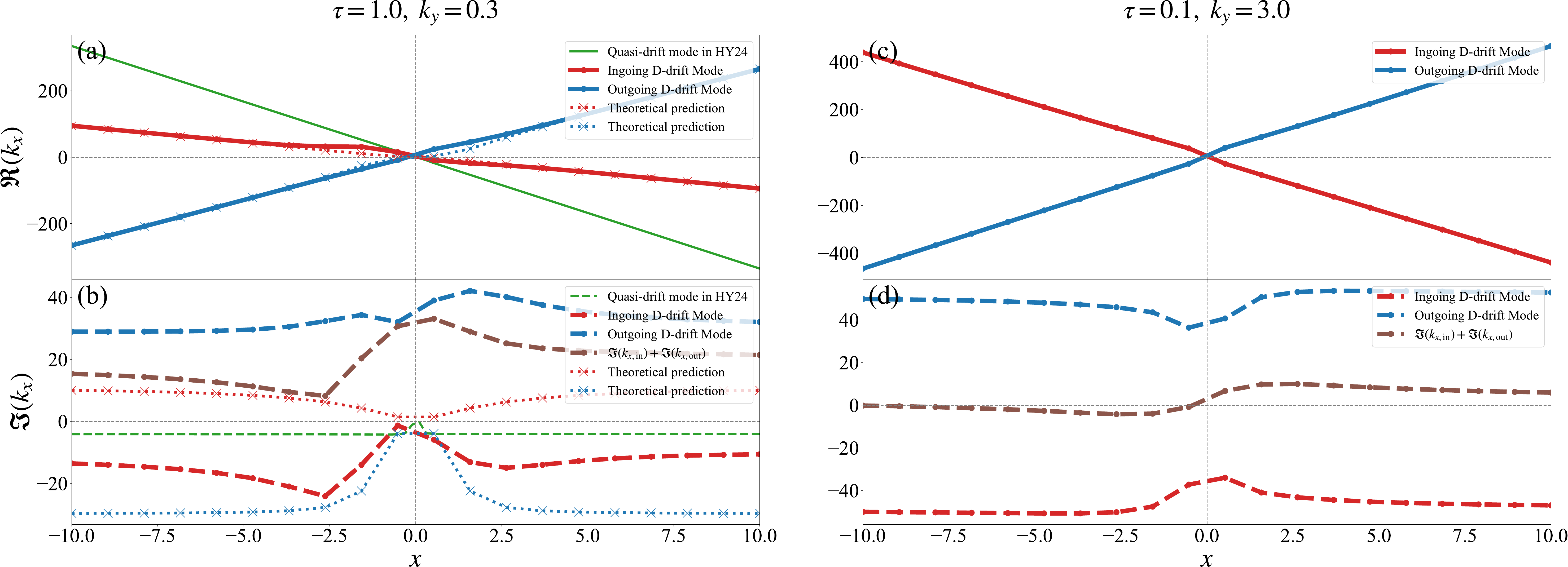}  
    \caption{The radial wavenumbers of D-drift modes with $\alpha = 1\times 10^{-3}$. The first row displays the real part of $k_x$, physically, the oscillating rate of waves. The second row displays the imaginary part of $k_x$, physically, the growth/decaying rate of waves. The first column is taken with $\tau=1.0, ~k_y=0.3$ for a illustration, while the second column is with $\tau=1.0, ~k_y= 3.0$ to show the importance of the outgoing D-drift mode. The results in \citetalias{Hou2024} are plotted as the green line for comparison. It can be found that the dust diffusion will smooth and damp the two D-drift modes much.} 
    \label{fig_wn}
\end{figure*}

In this section, we numerically calculate the radial wavenumbers, $k_x$, for the two predominant D-drift modes and compare them with both the analytical solution given by Equation \eqref{D-drift} and the quasi-drift mode described in \citetalias{Hou2024}.

Figure \ref{fig_wn} presents our results for $\alpha = 1 \times 10^{-3}$. The two columns correspond to the cases $\tau = 1.0, ~k_y = 0.3$ and $\tau = 0.1, ~k_y = 3.0$, respectively. The first row displays the real part of $k_x$, which physically represents the oscillation rate of the waves, while the second row shows the imaginary part of $k_x$, which physically represents the growth or decay rate of the waves. For an ingoing mode, $\Re{(k_x)}>0$ when $x<0$ and $\Re{(k_x)}<0$ when $x>0$. If the wave decays as it propagates, the condition $i k_x \Delta x < 0$ must hold, and with $\Delta x < 0$, this implies $\Im{(k_x)} < 0$. Conversely, an outgoing mode exhibits opposite characteristics, as shown in panels (a) and (b).

We observe smaller $\Re{(k_x)}$ and larger $|\Im{(k_x)}|$ in panels (a) and (b) compared to \citetalias{Hou2024} (green solid line). Both D-drift modes are significantly smoothed and dampened by dust diffusion. The dotted line represents our theoretical prediction from Equation \eqref{D-drift}, which aligns well with $\Re{(k_x)}$. However, there is some discrepancy in panel (b), likely due to neglected driving or damping terms. The dashed lines indicate that both D-drift modes are decaying as they propagate (at least away from the planet), with the ingoing D-drift mode exhibiting a slower decay rate than the outgoing mode. The brown dashed line represents the sum of $\Im{(k_x)}$ for both modes, which remains positive, suggesting the dominance of the ingoing D-drift mode. In this case, the outgoing D-drift mode is not clearly visible in either Figure \ref{fig_wave} or Figure \ref{fig_wake}.

However, as shown in panels (c) and (d), when $\tau = 0.1$, the two modes are comparable in both $\Re{(k_x)}$ and $\Im{(k_x)}$. In fact, the outgoing mode significantly contributes to the negative torque, as will be discussed in the next section.

\section{Disk Torque} \label{torque}
In this section, we calculate the torque exerted by the disk on the planet, with contributions from both dust and gas, using Equations(26)-(28) in \citetalias{Hou2024}. To ensure sufficient convergence and computational efficiency, we integrate them with $k_y$ ranging from $0.01$ to $15$, logarithmically discretized into 320 points (see the discussion in Section \ref{torque_distribution}). For convenience, we define $\Gamma_0 = q^2 h_p^{-3} \Sigma_{p} r_p^4 \Omega_{p}^2$. The total torque includes components of LT, CT, and ST. However, due to the complexities discussed in \citetalias{Hou2024}, it is challenging to fully disentangle these contributions. Therefore, we provide only a qualitative analysis of these torque components in the following sections.

\subsection{Torque Distribution} \label{torque_distribution}
\begin{figure*}[htbp]
    \centering
    \includegraphics[width=1.0\textwidth]{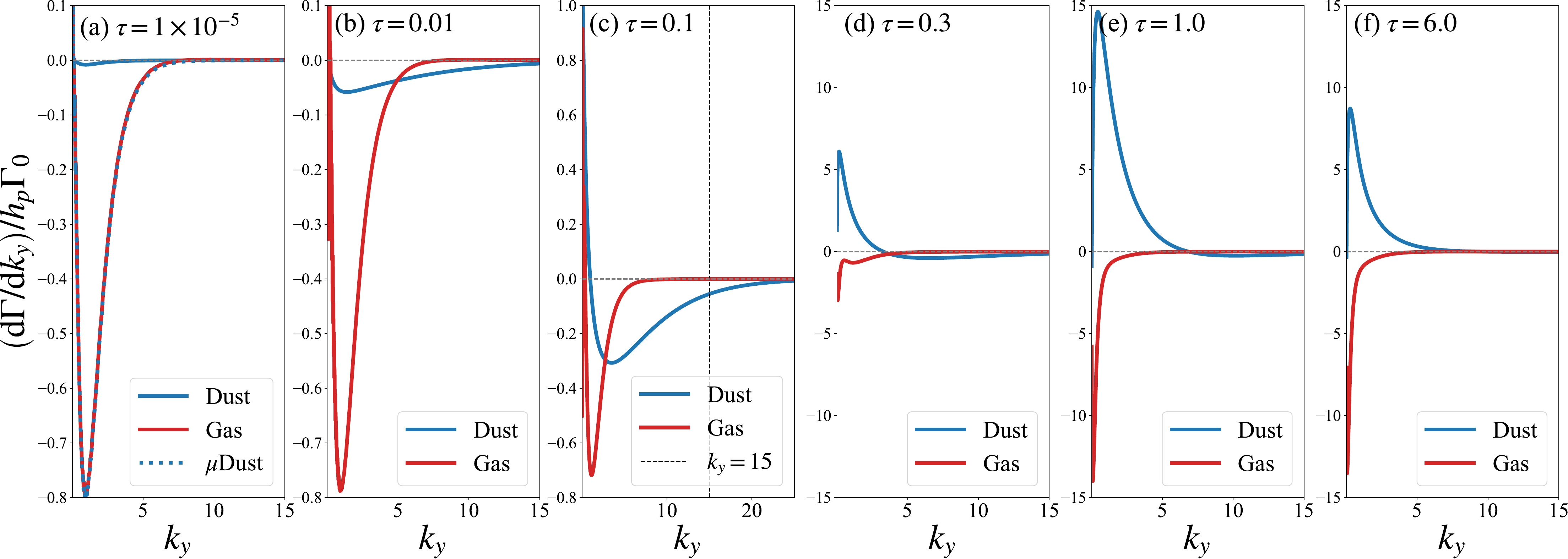}  
    \caption{The variation of dusty and gaseous torque with $k_y$. Five panels are given with different $\tau$ values, while they all have $\alpha = 1\times 10^{-3}$. If $\tau$ is small enough, the dusty behaviors are totally controlled by the gas, same for torque distribution as panel (a) shows. With $\tau$ increasing, the negative ST appears under the control of the gas as panel $(b)$ and $(c)$ shows. With much large $\tau$, the positive dusty torque mainly from small $k_y$ modes dominates, which is far lager than the negative gaseous torque as panel (d)-(f) show.}
    \label{torque_ky}
\end{figure*}
Figure \ref{torque_ky} displays the dusty and gaseous torques on the planet per $k_y$, i.e., $\mathrm{d}\Gamma / \mathrm{d} k_y$, with $\alpha= 1\times 10^{-3}$ and five different values of $\tau$. When $\tau = 1\times 10^{-5}$, as shown in panel (a), we plot $\mathrm{d}\Gamma / \mathrm{d} k_y$ for dust and gas, as well as $ \left( \mathrm{d}\Gamma_d / \mathrm{d} k_y\right) / \mu$ labeled by ``$\mu$Dust'', where the latter fits well with $\mathrm{d}\Gamma_g / \mathrm{d} k_y$. This fit implies consistent behaviors between dust and gas, while the former is entirely controlled by the latter. As $\tau$ increases, ST appears but is also influenced of gas, leading to a larger negative torque, as depicted in panels (b) and (c). Only positive torque appears at small $k_y$. In this scenario, the outgoing D-drift mode contributes significantly to the negative torque. In panel (c), we observe that the negative dusty torque reaches its maximum when $k_y \sim 3$, which explains our choice of $\tau = 0.1$ and $k_y = 3.0$ in Figure \ref{fig_wn} for illustration. Simultaneously, we note that the dusty torque per $k_y$ has not converged up to $k_y = 15$, while $k_y = 25$ seems to have achieved good convergence. In fact, the final integrated dusty torque differs little, less than $1\%$. For improved computational efficiency and convenient comparisons with other cases, we select a cut-off at $k_y = 15$ when $\tau \sim 0.1$. As $\tau$ increases, the dusty behavior gradually escapes the control of the gas, and ST contributes a large positive torque, primarily from the modes with small $k_y$. The torque distribution peaks when $\tau \sim 1$ and decreases at larger values of $\tau$, as shown in panels (d)-(f).
\begin{figure*}[htbp]
    \centering
    \includegraphics[width=1.0\textwidth]{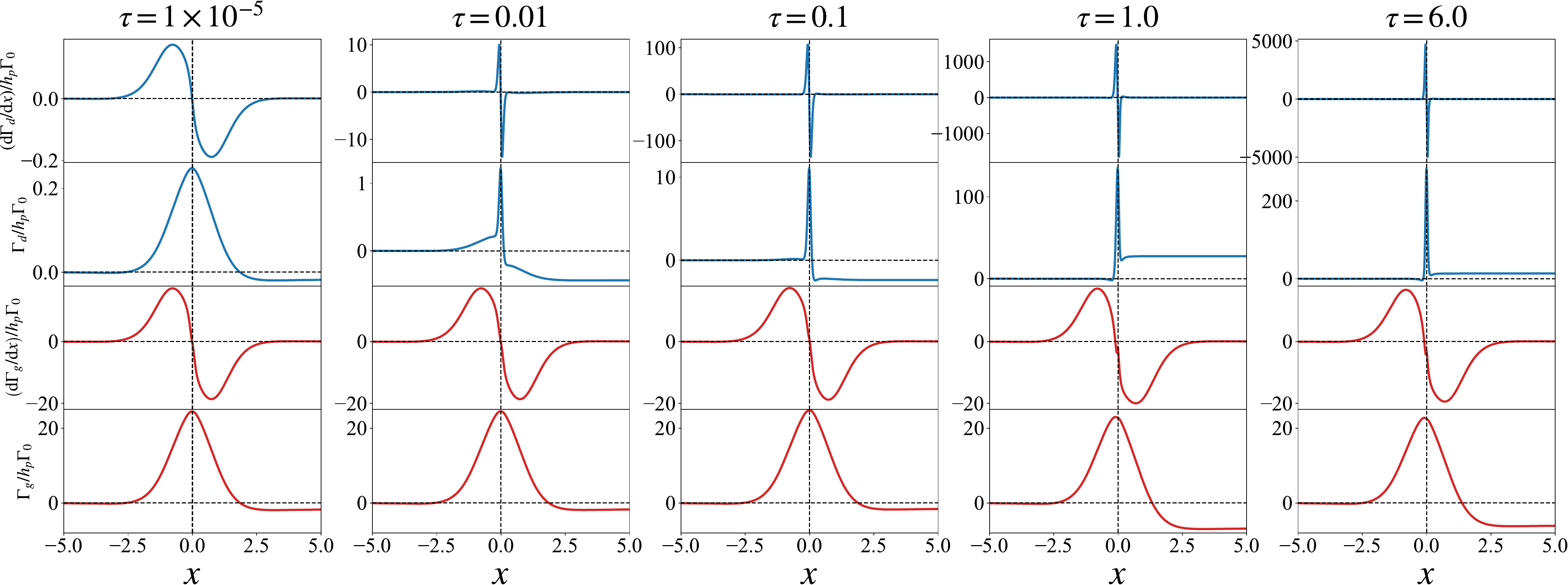}  
    \caption{Torque densities $\mathrm{d}\Gamma / \mathrm{d} x$ and integrated torque from the inner boundary to $x$ with $\alpha = 1\times 10^{-3}$. From the top row to the bottom row, they are dusty torque density, integrated dusty torque, gaseous torque density and integrated gaseous torque, respectively. From the leftmost column to the rightmost column, they correspond to the results of $\tau=1\times10^{-5}$, $\tau=0.01$, $~\tau=0.1$, $\tau=1.0$ and $\tau=6.0$. }
    \label{torque_density}
\end{figure*}

In Figure \ref{torque_density}, we plot the torque densities $\mathrm{d}\Gamma / \mathrm{d} x$ and the integrated torques with $\alpha = 1 \times 10^{-3}$ for different values of $\tau$. Note that the sign is opposite to some previous studies \citep[e.g.,][]{Rafikov2012,Fairbairn2024}, as they plot the torque exerted by planets on disks. The first row shows the dusty torque density, while the second row displays the integrated dusty torque from the inner boundary to $x$, exhibiting the same trend as shown in Figure \ref{torque_ky}. The third and fourth rows represent the gaseous torque density and integrated values, respectively. 

When $\tau = 1 \times 10^{-5}$, the dusty torque density is completely controlled by gas. As $\tau$ increases to $0.01$, the dusty torque density near the planet increases and dominates over that at larger $x$, indicating that ST becomes more significant than LT. The peak value of the dusty torque density continues to rise, while the integrated value reaches its maximum when $\tau \sim 1$. For the gaseous torque, the increased negativity is attributed to CT. An increased positive $u_{gx}$ causes gas particles to pass the planet more quickly, contributing additional AM and resulting in a larger negative CT \citep[e.g.][]{Masset2003,Ogilvie2006,Paardekooper2014,Wu2025}. Additionally, there are no distinct effects of dust diffusion on the gaseous torque distribution.

\subsection{Torque Dependence} \label{tau}
\begin{figure*}[htbp]
    \centering
    \includegraphics[width=\textwidth]{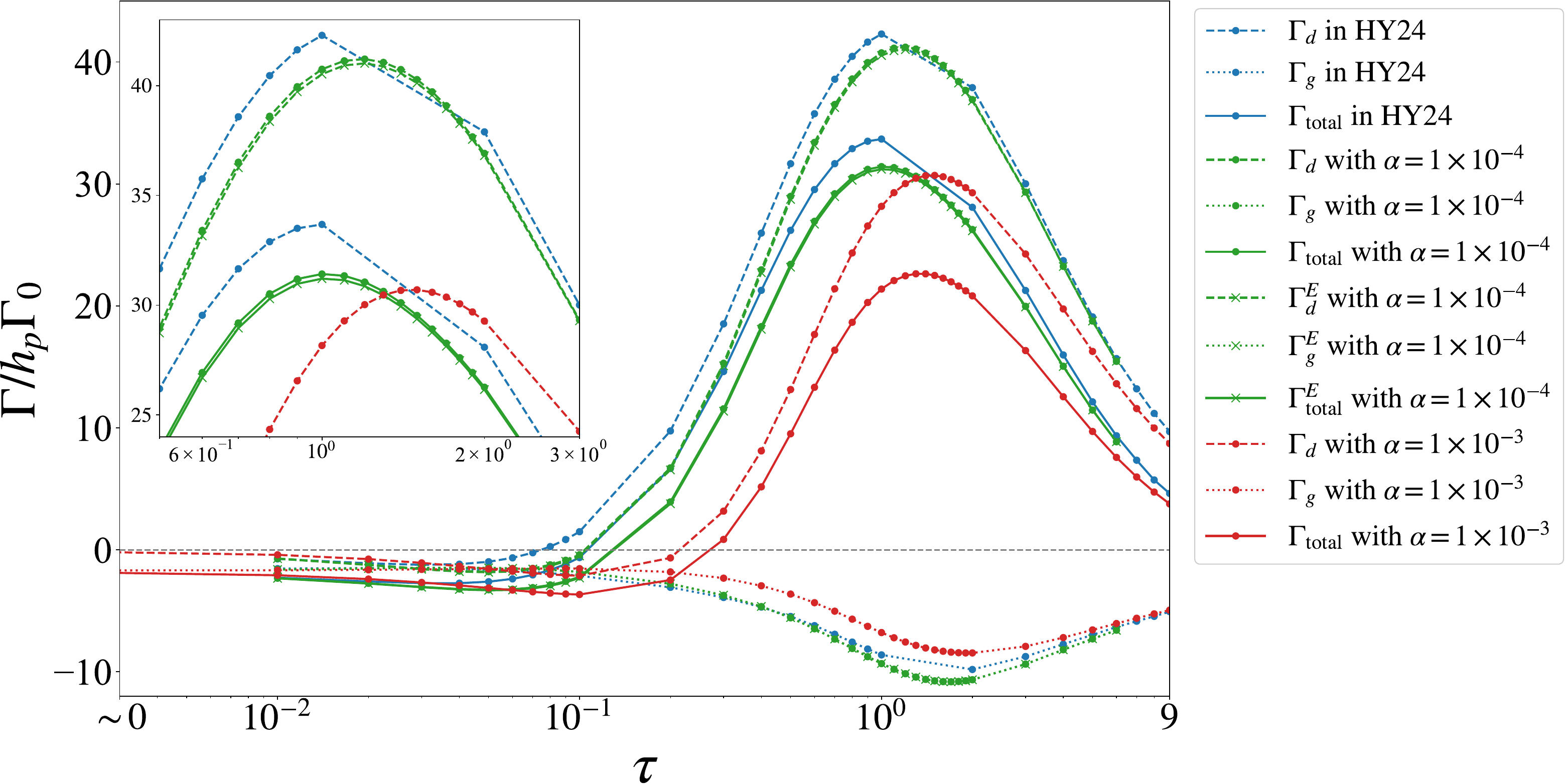}  
    \caption{Disk torque on planets with different $\tau$ and $\alpha$. The dashed, dotted and solid lines correspond to dusty, gaseous and total torque, respectively. The blue, green and red lines corresponds to the results of \citetalias{Hou2024}, with $\alpha= 1\times 10^{-4}$ and $\alpha= 1\times 10^{-3}$, respectively. The solid dots and cross in those lines correspond to our calculated results. The data denoted by cross and labeled by superscript ``$E$'' is from solving a new set of perturbed equation within the Epstein regime. A zoomed-in view is shown in the top-left panel, focusing on the region around $\tau = 1.0$.}
    \label{fig_torque}
\end{figure*}
We conduct a parameter survey for disk torque with different values of $\tau$ and $\alpha$, as shown in Figure \ref{fig_torque}. The dashed, dotted, and solid lines correspond to the dusty, gaseous, and total torques, respectively. The blue, green, and red lines represent the results from \citetalias{Hou2024}, with $\alpha = 1 \times 10^{-4}$ and $\alpha = 1 \times 10^{-3}$, respectively. The solid dots and crosses on these lines correspond to our calculated results. The data denoted by crosses and labeled with the superscript ``$E$'' comes from solving the set of perturbed equations within the Epstein regime (see Appendix \ref{appd_Epstein}). Notably, the torque results do not differ from those obtained in the Stokes regime. As shown in the figure, the crosses are nearly coincident with the dotted lines when $\alpha = 1 \times 10^{-4}$. A zoomed-in view is shown in the top-left panel, focusing on the region around $\tau = 1.0$, which gives a clear comparison.

As $\tau \rightarrow 0$, the dusty behavior is entirely controlled by the gas, and Figure \ref{fig_torque} demonstrates that $\Gamma_d/\Gamma_g = \mu$. As discussed in \citetalias{Hou2024}, the strength of ST is weak when $\tau$ is small. In this case, we observe that the dusty torque shows little variation with different values of $\alpha$. When $\tau \sim 1$, ST becomes predominant, and the total torque becomes positive and very large. However, the figure indicates that the dusty torque decreases as $\alpha$ increases. In addition to the suppression of the radial length scale previously demonstrated, we conclude that dust diffusion also restrains ST.

For the gaseous torque, we find that its negative value decreases slightly as well. Dust diffusion does not directly affect gaseous waves but does influence them through the dusty D-drift modes. These modes largely overlap with the planetary corotation region, and their large amplitude induces feedback on gaseous CT. We refer to this as the decrease of ST caused by dust diffusion, which results in a reduction in gaseous CT. However, we do not focus on this aspect here, and further confirmation might be needed.

Another important feature to note is that when $\tau \sim 0.1$, the negative dusty torque is larger compared to the results presented in \citetalias{Hou2024}. The underlying mechanism has been clarified previously: the torque contribution from the outgoing D-drift mode. Due to this contribution, the turning point for the dusty torque changes from negative to positive values, altering the direction of planetary migration. From Figure \ref{fig_torque}, we observe that the zero-torque turning points for the total torque shift from $\tau \simeq 0.1$ to $\tau \simeq 0.3$, which is not distinctly observable in a weak or moderate turbulence disk.

\subsection{Other Dependencies}
We also conducted additional dependence calculations for disk torque. Compared to \citetalias{Hou2024}, dust diffusion facilitates exploration of the system due to the shortened length scale of the D-drift mode.

\subsubsection{Dust mass fraction} \label{fraction}

\begin{figure}[htbp]
    \centering
    \includegraphics[width=1.0\columnwidth]{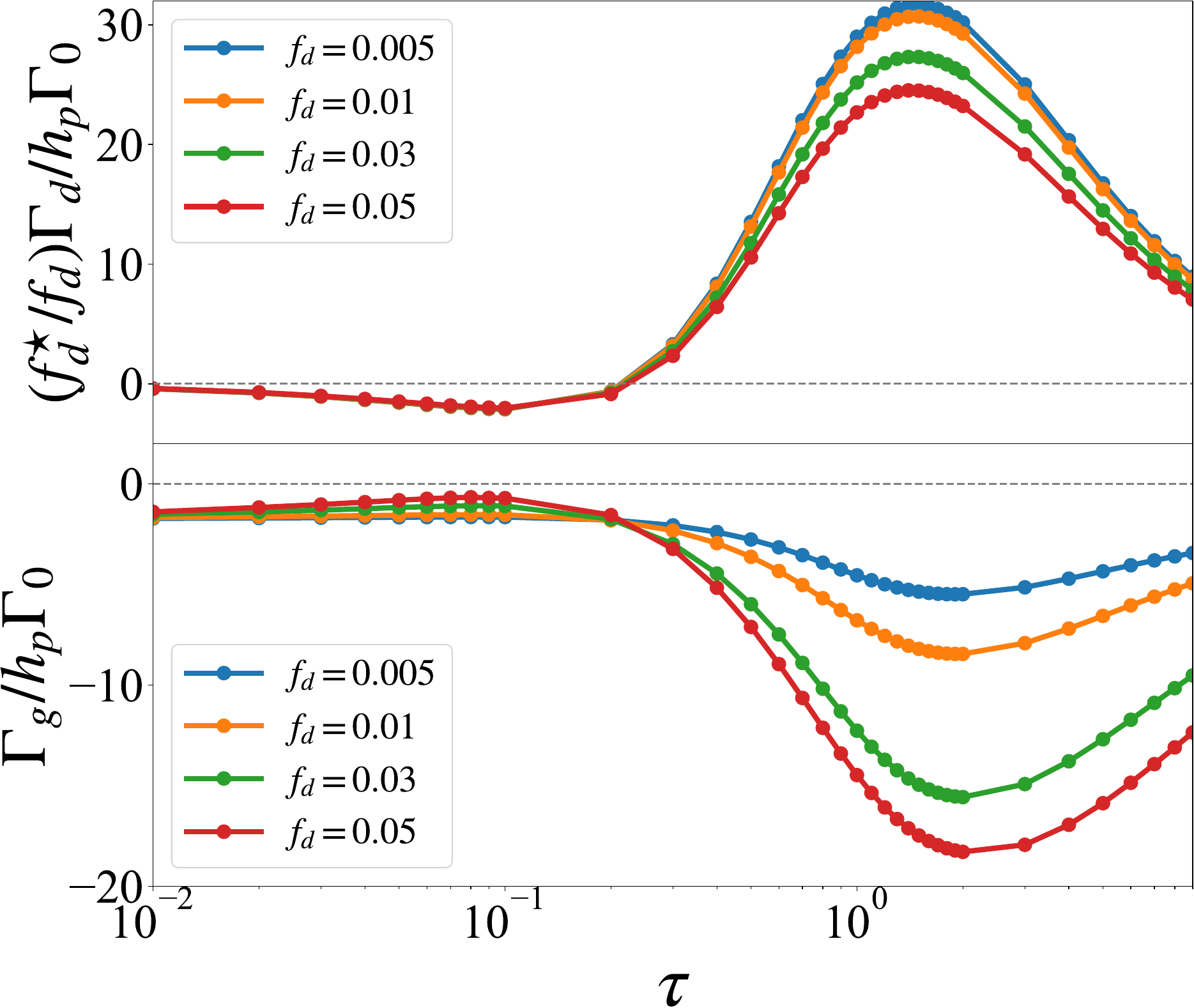}  
    \caption{The dependence of disk torque on dust mass fraction with $\alpha = 1 \times 10^{-3}$. Four lines with different colors represent the results with different $f_d$. The upper panel displays the dusty torque normalized by $(f_d/f_d^{\star}) h_p \Gamma_0$, where $f_d^{\star} = 0.01$. The lower panel displays the gaseous torque normalized by $h_p \Gamma_0$. }
    \label{fig_fraction}
\end{figure}

Figure \ref{fig_fraction} shows the dependence of disk torque on dust mass fraction with $\alpha = 1 \times 10^{-3}$. Four lines in different colors represent the results for various $f_d$. The upper panel displays the dusty torque normalized by $\left(\frac{f_d}{f_d^{\star}}\right) h_p \Gamma_0$, where $f_d^{\star} = 0.01$. In our range of $f_d$, a larger $f_d$ has little effect on dusty NSH velocities (see Equations \eqref{NSH_1}-\eqref{NSH_2}), thus the D-drift modes remain largely unaffected. As a result, the dusty torque is almost directly proportional to the dust mass fraction. For the gas, the gaseous radial velocity increases with an increasing $f_d$ (see Equation \eqref{NSH_3}). Therefore, the negative CT contribution is amplified by a larger $f_d$, which is most evident when $\tau \simeq 1$. Consequently, the total torque will be either amplified or reduced by the dust mass fraction. The zero-torque turning point remains on the order of $\tau \sim 10^{-1}$ (at least within the $f_d$ range we investigated).

\subsubsection{Scale height} \label{sh}
\begin{figure}[htbp]
    \centering
    \includegraphics[width=1.0\columnwidth]{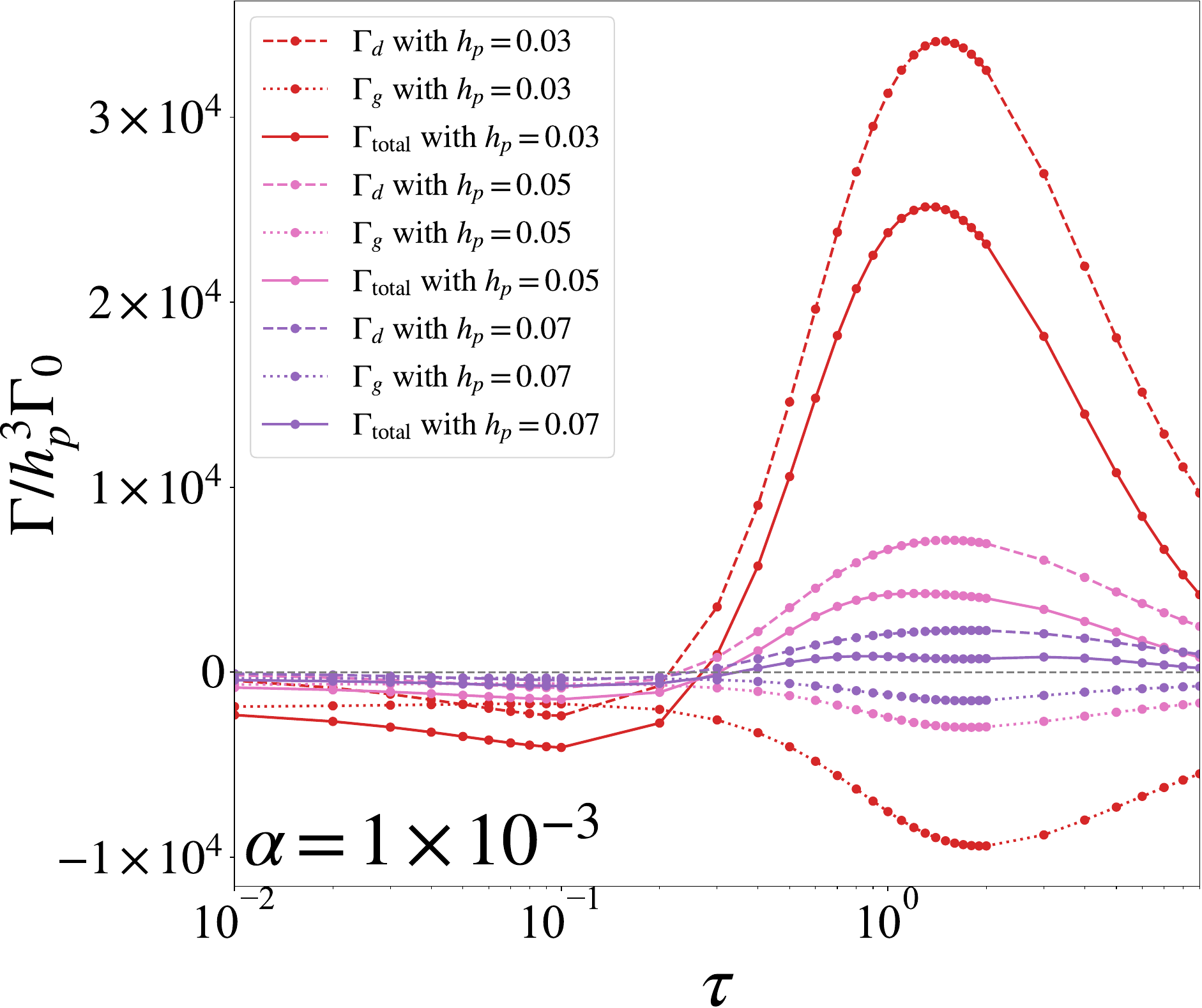}  
    \caption{The dependence of disk torques on scale height with $\alpha=1\times 10^{-3}$. Dashed, dotted and solid lines correspond to dust, gas and total torques, respectively. Three different colors correspond to different values of scale height.}
    \label{fig_sh}
\end{figure}
Equation \eqref{NSH_1} indicates that dusty drift velocity increases with an increasing scale height, which will result in an increased ST when normalized by $\Gamma_0$. However, both $\Gamma_0$ and the final planetary migration timescale are also related to scale height. Therefore, we plot disk torques normalized by $h_p^3 \Gamma_0 = q^2 \Sigma_p r_p^4 \Omega_p^2$, which directly reflects the speed of migration. The results are displayed in Figure \ref{fig_sh}. Dust diffusion is set with $\alpha = 1 \times 10^{-3}$. Lines in the figure with different colors represent different scale heights, as indicated in the legend. It is evident that the migration speed decreases with a larger $h_p$. Additionally, the zero-torque turning point remains nearly unchanged (at least within our investigated parameter space).

\section{Discussions} \label{discussion}
\subsection{Planetary Migration} \label{migration}
Using the above results, we can estimate the timescale of planetary migration. With a typical dust mass ratio of $f_d = 0.01$, we find that the typical timescale for an Earth-sized planet is not smaller than that given by \citetalias{Hou2024}, i.e., $10^5$ years (whether inward or outward). For a larger dust mass ratio, protoplanets might migrate faster, or complex fluid motion may occur \citep{Hsieh2020}. Although planetary migration depends on many parameters in a dust-gas coupled protoplanetary disk (PPD) as we calculated, the zero-torque turning point remains almost unchanged, which determines the direction of migration. The corresponding critical stopping time is $\tau = 0.1$. For a single size of dust, we can estimate the corresponding critical dust size $a_{\text{cri}}$. Although dust size has a distribution in practice, the width of the streaming torque density is quite narrow, so only dust sizes near the position of the planet matter. The expression for the stopping time has been given by Equation \eqref{regime}. For a low-mass protoplanet embedded in the minimum mass solar nebula, where $\Sigma_g \simeq 150 \left(\frac{r}{ 5 \rm{AU}} \right)^{-1.5} ~\rm{g \cdot cm^{-2}}$ \citep{Hayashi1981} and $\rho_{\rm{int}} \simeq 1.4 ~\rm{g \cdot cm^{-3}}$, we let $\rho_g = \Sigma_g / \left( \sqrt{2\pi} H_p \right)$, which corresponds to the gas density at midplane. Adopting $\sigma_g \simeq 2\times 10^{-15} ~ \rm{cm}^2$ as the cross-section for the collisions of molecular hydrogen \citep{Armitage2020}, we can finally obtain
\begin{equation} \label{estimate}
    a_{\text{cri}} \simeq 
    \begin{cases}
        \displaystyle 48 ~ \text{cm}, & \text{if } r < 1.37 \text{AU}, \\[10pt]
        \displaystyle 48 \left(\frac{r}{ 1.37 \rm{AU}} \right)^{-1.5} ~ \text{cm},  & \text{if } r \geq 1.37 \text{AU}.
    \end{cases}
\end{equation}
This indicates that the Stokes regime applies in the inner region of PPDs, where $a_{\text{cri}}$ is constant. In the outer region, the Epstein regime applies, and the critical dust size corresponds to large pebbles, which decreases as the orbital radius increases. To achieve outward migration, we require $\tau > 0.1$. The above estimations imply that fast coagulation of dust grains will help planets survive, while gas turbulence will hinder the rapid dust coagulation. Therefore, our results suggest that weak gas turbulence makes it easier for planets to survive. We believe that Equation \eqref{estimate} will be meaningful in comprehensive scenario studies that combine planetary migration and planet formation through pebble accretion \citep[e.g.,][]{Ormel2010, Lambrechts2012}.

\subsection{Dust Diffusion Prescription} \label{model}
\begin{figure*}[htbp]
    \centering
    \includegraphics[width=1.0\textwidth]{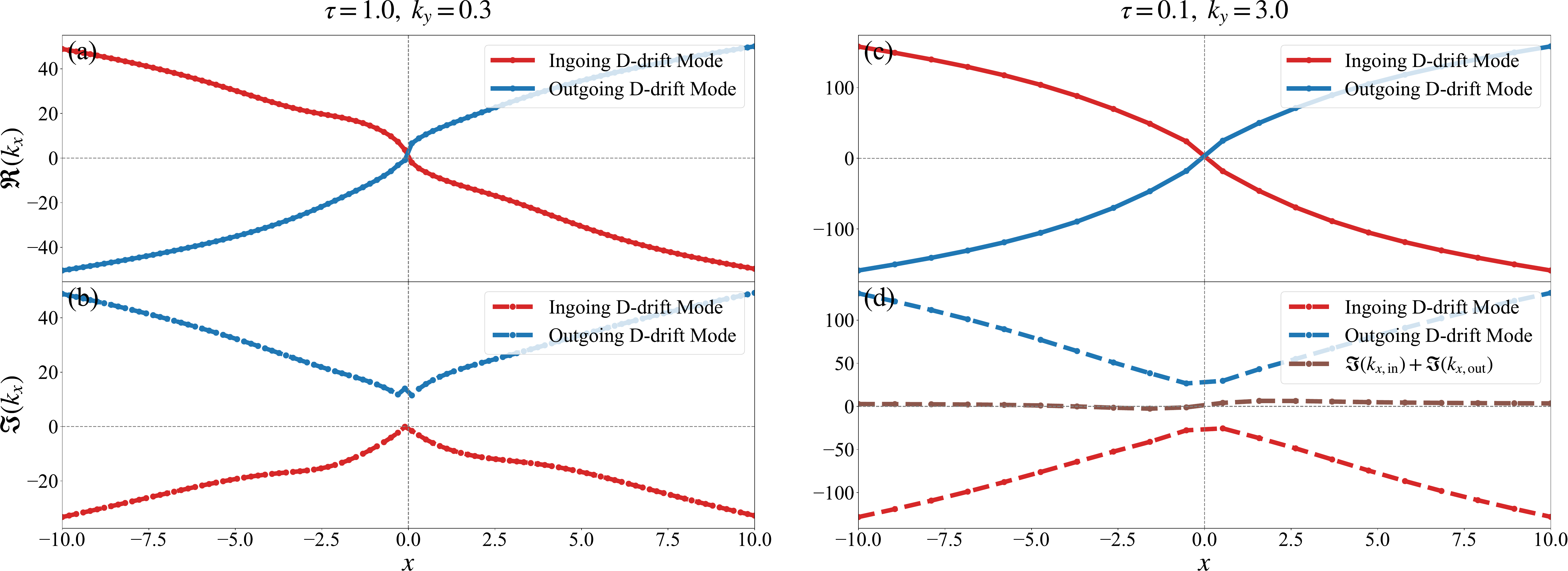}  
    \caption{Using Equation \eqref{eq_dust_in1}-\eqref{eq_dust_in2} but not \eqref{eq_dust_1}-\eqref{eq_dust_2} for dust diffusion prescription, the radial wavenumbers of D-drift modes with $\alpha = 1\times 10^{-3}$. Similar to Figure \ref{fig_wn}, the first row displays the real part of $k_x$, physically, the oscillating rate of waves. The second row displays the imaginary part of $k_x$, physically, the growth/decaying rate of waves. The first column is taken with $\tau=1.0, ~k_y=0.3$ for a illustration, while the second column is with $\tau=1.0, ~k_y=3.0$ to show the importance of the outgoing D-drift mode.}
    \label{fig_wn_new}
\end{figure*}
In this research, we use the formulation of dust diffusion given by \cite{Klahr2021} and also describe the formulation used in previous studies (see Equations \eqref{eq_dust_in1} and \eqref{eq_dust_in2}). In this subsection, we plot the radial wavenumbers of the D-drift modes using the latter prescription for dust diffusion. Although it is self-inconsistent, the added physical ingredient is the same.

The results are displayed in Figure \ref{fig_wn_new}, which is similar to Figure \ref{fig_wn}. We identify two predominant D-drift modes. In the first column, when $\tau = 1.0$ and $k_y = 0.3$, the D-drift modes exhibit a smaller oscillation rate but similar decay rates compared to those in Figure \ref{fig_wn}. In the second column, when $\tau = 0.1$ and $k_y = 3.0$, both the oscillation and decay rates are larger than those in Figure \ref{fig_wn}. However, the decay rates of the two D-drift modes remain comparable, indicating a larger negative torque at $\tau \sim 1$, consistent with our previous findings.

Through the above comparison, we conclude that our inferences are correct (at least qualitatively). Future studies can also conduct tests using other dust diffusion prescriptions \citep{Tominaga2019, HPH2022}.

\subsection{The Choice of Smoothing Length} \label{soft}
An important consideration in torque calculations is the choice of smoothing length for the planetary potential, as it directly influences the final torque estimates. In 2D simulations, the smoothing length is introduced to approximate the effects of three-dimensional (3D) geometry and to mitigate the singularity at the planet’s location.

Previous studies focusing on pure gas PPDs have employed smoothing lengths typically ranging from $0.4 H_p$ to $0.7 H_p$. However, a comparison of 3D and 2D torque calculations by \cite{Tanaka2024} highlights the difficulty of capturing all the dependencies of 3D torques using 2D torque formulas with a single smoothing length. This limitation arises because the smoothing length modifies the shape of the planetary potential, affecting different components of the torque asymmetrically. For example, LT and CT respond to distinct regions of the potential, which are influenced by variations in pressure gradients, density gradients, and other factors.

When $\tau$ is sufficiently small, the gaseous torque of approximately $-2 h_p \Gamma_0$ aligns well with the value of $\sim -2.3 h_p \Gamma_0$ reported in \cite{Tanaka2002}, which is predominantly attributed to LT. In our model, which considers only the pressure gradient, the top-left panel of Figure 5 in \cite{Tanaka2024} demonstrates that our chosen smoothing length of $1/8 H_p$ is sufficiently small to recover the gaseous LT accurately. It is worth noting that the value of $-1.436 h_p \Gamma_0$ in that panel also includes the CT contribution from the curvature term, which is absent in our study. For larger values of $\tau$, the gaseous CT becomes more significant due to the influence of radial flow. In such cases, a larger smoothing length is more appropriate, suggesting that our current choice might overestimate the negative gaseous torque in these scenarios.

In a dust-gas coupled PPD, the calculation of ST becomes more complex. The 3D structure of the dust differs from that of the gas, as the vertical distribution of dust is not supported by pressure but instead by turbulent agitation from the gas. The effective dust scale height can be expressed as $H_d = H_p \sqrt{\frac{\alpha}{\tau + \alpha}}$ \citep{Dubrulle1995}. Only when $\tau$ is extremely small does the dust share the same vertical structure as the gas. For general values of $\tau$, with typical values of $\alpha$ ranging from $10^{-5}$ to $10^{-3}$, the effective dust scale height is approximately $H_d \simeq \sqrt{\alpha / \tau} H_p$, resulting in $H_d \sim 10^{-2} - 10^{-1} H_p$. This suggests that a smoothing length of $1/8 H_p$ might be large, potentially leading to an underestimation of ST. A smaller smoothing length should therefore be used when $\alpha$ is smaller, indicating that gas turbulence not only suppresses the quasi-drift mode but also moderates the magnitude of ST through affecting the planetary potential.

Further investigation using 3D semi-analytical calculations, such as decompositions of the vertical structure with Hermite polynomials \citep{Takeuchi1998,Tanaka2002,Tanaka2024}, may help identify a more suitable smoothing length for ST. Alternatively, \cite{Brown2024} proposed an approach that averages the planetary potential in the vertical direction, showing improved accuracy over smoothing methods in pure gas 2D PPD calculations. However, this method retains the singularity in the planetary potential, raising concerns about its applicability for calculating CT magnified by radial flow and for estimating ST. Further discussions are needed to assess its reliability in these contexts.}

In realistic PPDs, the dynamics of dust near planets are highly complex, adding further challenges to determining an appropriate smoothing length. For instance, dust accretion onto the planet may amplify the positive dusty torque \citep{Regaly2020, Chrenko2024}. 2D hydrodynamic simulations by \cite{Llambay2018} and \cite{Chrenko2024} adopt a smoothing length equivalent to the Hill radius, defined as:
\begin{equation}
    R_{\rm{H}}= r_p \left( \frac{M_{p}}{3M_{\star}} \right)^{1/3} = H_{p} \left( \frac{M_{p}}{3M_{\rm{th}}} \right)^{1/3}.
\end{equation}
Assuming $M_{\rm{th}}$ to be comparable to Jupiter’s mass and $M_p$ to range from 1 to 10 Earth masses, the Hill radius is approximately $R_{\rm{H}} \simeq 0.1 - 0.2 H_p$, which aligns well with our choice of a smoothing length of $1/8 H_p$.

In conclusion, adopting a constant smoothing length of $ 1/8 H_p$ is reasonable for now, which can also isolate the effects of dust diffusion on the quasi-drift mode. However, it is important to note that the magnitude of ST is highly sensitive to the planetary potential and the complex environmental conditions near the planet. Future studies should focus on 3D calculations, whether through semi-analytical approaches or hydrodynamic simulations, to address these intricacies. Special attention must be given to the immediate vicinity of planets, which poses significant challenges. Nevertheless, robust 3D foundations for the selection of smoothing length can be established through such efforts, paving the way for more accurate torque calculations.

\section{Summary and Conclusions} \label{conclusion}
In the paper, we investigate the effects of dust diffusion on ST which is proposed to address the fast and inward type-I migration problem for low-mass protoplanets. A linear analysis is conducted based on NSH equilibrium, beyond which ST can also exist. We utilize the formulation of dust diffusion presented by \cite{Klahr2021}, while we also make a comparison between different formulations at last. Our findings indicate that they are qualitatively similar. We hope that future studies, aided by hydrodynamic simulations, will further demonstrate this point.

We first derive the dispersion relation for the two-fluid system using the WKB approximation and find that six modes exist. Both gas and dust have two DWs with different propagation directions, where the gaseous DWs are restored by pressure and the dusty DWs are driven by the aerodynamic drag. We also identify two predominant D-drift modes with different directions. In fact, these modes are split from the quasi-drift mode due to dust diffusion. We can easily recover the free or quasi-drift mode by eliminating dust diffusion. The two D-drift modes exhibit smaller radial wavenumbers and larger damping compared to the quasi-drift mode. This indicates dust diffusion can lengthen and dampen the extremely short structure of the quasi-drift mode, leading to smoothed dusty wakes (see Figure \ref{fig_wake}). The ingoing D-drift mode is very similar to the quasi-drift mode but contributes a smaller positive torque to the planet in most cases. The outgoing mode dominates only when $\tau \sim 0.1$, which alters the zero-torque turning point. However, this change is minimal, and the stopping time remains on the order of $\tau \sim 0.1$. In conclusion, dust diffusion generally restrains ST and is detrimental to the survival of protoplanets.
 
We also present the torque distributions, including disk torque per $k_y$ and torque density. When $\tau \rightarrow 0$, the distributions of dusty torque are entirely controlled by the gas and are directly proportional to the dust-gas mass ratio. In this case, only CT and LT exists. As $\tau$ increases to relatively small values ($\lesssim 0.1$), negative ST under the control of gas appears. When $\tau$ is large enough ($\gtrsim 0.1$), a large and positive ST concentrated at small $k_y$ becomes dominant. Concurrently, the gaseous CT changes a lot due to variations in radial velocity.

We then conduct some parametric surveys with moderate dust diffusion, exploring the dependence of disk torque on the Epstein regime, dust mass fraction and scale height. Compared to \citetalias{Hou2024}, dust diffusion simplifies our numerical calculations and enables us to explore a broader parameter space due to the smoothness of the quasi-drift mode. In future studies, hydrodynamic simulations for ST will benefit from incorporating dust diffusion to achieve better convergence. In the Epstein regime, we perturb the basic equations with $\Sigma_g t_s = \text{const}$. However, the final gaseous and dusty torques remain the same as those in the Stokes regime. It is noteworthy that in both regimes, we assume that the aerodynamic drag is proportional to the first power of the relative velocity between dust and gas, while a different power-law dependence might be used in the Stokes regime \citep{Weidenschilling1977, Armitage2020}.

Within a limited dust mass fraction range ($0.005-0.05$), we find that  the dusty torque is almost directly proportional to dust mass fraction. In contrast, the gaseous torque, particularly CT, varies significantly due to changes in radial velocity with the dust mass fraction. Regarding the scale height ($0.03-0.07$), both gaseous and dusty torques decrease with its increasing, which will slower planetary migration (whether inward or outward). However, in all the cases we considered, the zero-torque turning point changes little, and the corresponding stopping time consistently remains on the order of $\tau \sim 0.1$, delineating the boundary for planetary migration directions. The corresponding dust grains (see Equation \eqref{estimate}) are larger pebbles, implying that fast dust coagulation in weakly turbulent PPDs will aid in planetary survival. These results will be significant for future studies that integrate planetary migration and core accretion.

\section*{Acknowledgment}
We thank the anonymous referee for the helpful comments and suggestions that clarify and improve the manuscript. Valuable discussions with Shu-ichiro Inutsuka and Ryosuke T. Tominaga are also highly appreciated. This work has been supported by the National SKA Program of China (Grant No. 2022SKA0120101) and National Key R\&D Program of China (No. 2020YFC2201200) and the science research grants from the China Manned Space Project (No. CMS-CSST-2021-B09, CMS-CSST-2021-B12, and CMSCSST-2021-A10) and opening fund of State Key Laboratory of Lunar and Planetary Sciences (Macau University of Science and Technology) (Macau FDCT Grant No. SKL-LPS(MUST)-2021-2023). C.Y. has been supported by the National Natural Science Foundation of China (grants 11521303, 11733010, 11873103, and 12373071).

\vspace{5mm}
\software{NUMPY \citep{Harris2020}, MATPLOTLIB \citep{Hunter2007}}
\appendix
\section{Linear Perturbation Equations for Different Regimes} \label{appd_Epstein}
For aerodynamic drag between dust and gas, stopping time reads
\begin{equation} \label{regime}
    t_s =
\begin{cases}
    \displaystyle \sqrt{\frac{\pi}{8}} \frac{\rho_{\text{int}} a}{\rho_g c_s}, & \text{if } a \leq \frac{9}{4} l_m \quad \left( \text{Epstein regime} \right), \\[10pt]
    \displaystyle\sqrt{\frac{\pi}{8}} \frac{4 \rho_{\text{int}} a^2}{9 \rho_g c_s l_{\text{mfp}}}, & \text{if } a > \frac{9}{4} l_m \quad \left( \text{Stokes regime} \right).
\end{cases}
\end{equation}
where $\rho_{\rm{int}}$ is the internal material density of dust grains and $a$ is the size of dust particles. $l_{\text{mfp}} \propto 1/(\rho_g \sigma_g)$ is the mean-free path of gas molecules, where $\sigma_g$ is the cross-section for gaseous collisions. So, we let $ \Sigma_g t = \rm{const.}$ in Epstein regime and  $ t = \rm{ const.}$ in Stokes regime when perturbing Equation \eqref{eq_gas_2} and \eqref{eq_dust_2}. 
The set of perturbation equations in the Stokes regime read
\begin{gather}
    \left( i k_y \tilde{u}_{gy} + \tilde{u}_{gx} \frac{d}{d x} \right) \frac{\delta \tilde{\Sigma}_g}{f_g} + \frac{d}{d x} \delta \tilde{u}_{gx} + i k_y \delta \tilde{u}_{gy} = 0, \label{linear1}\\
    \left( \frac{\mu \tilde{w}_{s,x}}{\tau} + \frac{d}{dx} \right) \frac{\delta \tilde{\Sigma}_g}{f_g} + \left( \frac{ \mu}{\tau} + i k_y\delta \tilde{u}_{gy} + \tilde{u}_{gx}\frac{d}{dx} \right) \delta \tilde{u}_{gx} - 2 \delta \tilde{u}_{gy} - \frac{ \mu \tilde{w}_{s,x}}{\tau} \frac{\delta \tilde{\Sigma}_d}{f_d} - \frac{\mu }{\tau} \delta \tilde{u}_{dx}  = - \frac{\partial \phi_p}{\partial x}, \label{linear2} \\
    \left( \frac{\mu \tilde{w}_{s,y}}{\tau} + i k_y \right) \frac{\delta \tilde{\Sigma}_g}{f_g} + \frac{\delta \tilde{u}_{gx}}{2} + \left(\frac{\mu}{\tau} + i k_y \tilde{u}_{gy} + \tilde{u}_{gx} \frac{d}{d x} \right) \delta \tilde{u}_{gy} - \frac{\mu \tilde{w}_{s,y}}{\tau} \frac{\delta \tilde{\Sigma}_d}{f_d} - \frac{\mu}{\tau} \delta \tilde{u}_{dy} = - i k_y \phi_p, \label{linear3} \\
    \left( i k_y \tilde{u}_{dy} + \tilde{u}_{dx}\frac{d}{d x} \right) \frac{\delta \tilde{\Sigma}_d}{f_d} + \frac{d}{d x} \delta \tilde{u}_{dx} + i k_y \delta \tilde{u}_{dy} = 0, \label{linear4}\\
    - \frac{\alpha}{\tau} \frac{d }{dx} \left(\frac{\delta \tilde{\Sigma}_g}{f_g}\right) - \frac{1}{\tau} \delta \tilde{u}_{gx} + \frac{\alpha}{\tau} \frac{d }{dx} \left(\frac{\delta \tilde{\Sigma}_d}{f_d}\right) + \left( \frac{1}{\tau} + i k_y \tilde{u}_{dy} + \tilde{u}_{dx} \frac{d}{d x} \right) \delta \tilde{u}_{dx} - 2 \delta \tilde{u}_{dy} = - \frac{\partial \phi_p}{\partial x}, \label{linear5}\\
    - \frac{i k_y \alpha \delta \tilde{\Sigma}_g}{\tau f_g} - \frac{1}{\tau} \delta \tilde{u}_{gy} + \frac{i k_y \alpha \delta \tilde{\Sigma}_d}{\tau f_d }+ \frac{\delta \tilde{u}_{dx}}{2}
    + \left( \frac{1}{\tau} + i k_y \tilde{u}_{dy} + \tilde{u}_{dx}  \frac{d}{d x} \right) \delta \tilde{u}_{dy} = - ik_y \phi_p. \label{linear6}
\end{gather}
In the Epstein regime, Equation \eqref{linear2},  \eqref{linear3}, \eqref{linear5} and \eqref{linear6} are substituted by
\begin{gather}
    \frac{d}{dx} \frac{\delta \tilde{\Sigma}_g}{f_g} + \left( \frac{ \mu}{\tau} + i k_y\delta \tilde{u}_{gy} + \tilde{u}_{gx}\frac{d}{dx} \right) \delta \tilde{u}_{gx} - 2 \delta \tilde{u}_{gy} - \frac{ \mu \tilde{w}_{s,x}}{\tau} \frac{\delta \tilde{\Sigma}_d}{f_d} - \frac{\mu }{\tau} \delta \tilde{u}_{dx}  = - \frac{\partial \phi_p}{\partial x}, \label{e_linear2} \\
    \frac{ i k_y \delta \tilde{\Sigma}_g}{f_g} + \frac{\delta \tilde{u}_{gx}}{2} + \left(\frac{\mu}{\tau} + i k_y \tilde{u}_{gy} + \tilde{u}_{gx} \frac{d}{d x} \right) \delta \tilde{u}_{gy} - \frac{\mu \tilde{w}_{s,y}}{\tau} \frac{\delta \tilde{\Sigma}_d}{f_d} - \frac{\mu}{\tau} \delta \tilde{u}_{dy} = - i k_y \phi_p, \label{e_linear3} \\
    \left( \frac{ \tilde{w}_{s,x}}{\tau}  - \frac{\alpha}{\tau} \frac{d }{dx} \right) \frac{\delta \tilde{\Sigma}_g}{f_g} - \frac{1}{\tau} \delta \tilde{u}_{gx} + \frac{\alpha}{\tau} \frac{d }{dx} \left(\frac{\delta \tilde{\Sigma}_d}{f_d}\right) + \left( \frac{1}{\tau} + i k_y \tilde{u}_{dy} + \tilde{u}_{dx} \frac{d}{d x} \right) \delta \tilde{u}_{dx} - 2 \delta \tilde{u}_{dy} = - \frac{\partial \phi_p}{\partial x}, \label{e_linear5}\\
    \left( \frac{ \tilde{w}_{s,y}}{\tau} - \frac{i k_y \alpha }{\tau} \right)\frac{\delta \tilde{\Sigma}_g}{f_g} - \frac{1}{\tau} \delta \tilde{u}_{gy} + \frac{i k_y \alpha \delta \tilde{\Sigma}_d}{\tau f_d }+ \frac{\delta \tilde{u}_{dx}}{2} + \left( \frac{1}{\tau} + i k_y \tilde{u}_{dy} + \tilde{u}_{dx}  \frac{d}{d x} \right) \delta \tilde{u}_{dy} = - ik_y \phi_p. \label{e_linear6}
\end{gather}
The planetary potential, denoted as $ \phi_{p} $, are expressed using Bessel functions, same as that in \citetalias{Hou2024}.

\section{Dispersion Relation} \label{appd_DR}
With Fourier decomposition in $y$ direction, the perturbation reads
\begin{equation} \label{ansatz1}
    X_1 = \int_{- \infty}^{+ \infty} \delta X (x) \exp \left( ik_y y - i \tilde{\omega}t\right) d k_y, 
\end{equation}
where $\tilde{\omega}= \omega - k_y h_p^{-1} \Omega$ is the Doppler-shifted frequency of some certain mode. $\omega$ and $\Omega$ are the frequency of the mode and the rotation frequency of the coordinate, respectively. Adopting WKB approximation, we assume
\begin{equation} \label{ansatz2}
    \delta X(x) = A(x) \mathrm{exp}\left(i \int^{x} k_x(x^{\prime }) dx^{\prime } \right), 
\end{equation}
where $A$ and $k_x$ is the amplitude and radial wavenumber, correspondingly, both varying with $x$. For a particular $x$ (or $k_x$), combining ansatz \eqref{ansatz1}, \eqref{ansatz2} and perturbation equations, we obtain the equations for the eigenmodes, which in matrix read
\begin{equation}
    \begin{bmatrix}
        - i \tilde{\omega}_g^{\prime }  & i k_x & i k_y & 0 & 0 & 0 \\

        i k_x + \frac{\mu \tilde{w}_{s,x}}{\tau}  & - i \tilde{\omega}_g^{\prime } + \frac{\mu}{\tau} & - 2 & - \frac{\mu \tilde{w}_{s,x}}{\tau} & - \frac{\mu}{\tau} & 0 \\

        i k_y + \frac{\mu \tilde{w}_{s,y}}{\tau} & \frac{1}{2} & - i \tilde{\omega}_g^{\prime } + \frac{\mu}{\tau} & - \frac{\mu \tilde{w}_{s,y}}{\tau} & 0 & - \frac{\mu}{\tau} \\

        0 & 0 & 0 & - i \tilde{\omega}_d^{\prime } & i k_x & i k_y \\

        - \frac{\alpha i k_x}{\tau} & - \frac{1}{\tau} & 0 & \frac{\alpha i k_x}{\tau} & - i \tilde{\omega}_d^{\prime } + \frac{1}{\tau} & - 2 \\

        - \frac{ \alpha i k_y}{\tau} & 0 & -\frac{1}{\tau} & \frac{ \alpha i k_y}{\tau} & \frac{1}{2} & - i \tilde{\omega}_d^{\prime } + \frac{1}{\tau} & 

    \end{bmatrix}
    \begin{bmatrix}
        f_g^{-1} \delta \tilde{\Sigma}_{g}\\
        \delta \tilde{u}_{gx} \\
        \delta \tilde{u}_{gy} \\
        f_d^{-1} \delta \tilde{\Sigma}_{d}  \\
        \delta \tilde{u}_{dx} \\
        \delta \tilde{u}_{dy}
    \end{bmatrix} = 0,
\end{equation}
where $ \tilde{\omega}_{g/d}^{\prime } = \tilde{\omega}_{g/d} - k_x u_{gx/dx}$ and $ \tilde{\omega}_{g/d} = \tilde{\Omega}_{g/d}/\left( q h_p^{-3}\Omega_p \right)$. Note that the nonhomogeneous terms, i.e. the planetary potential, are discarded compared to Equation \eqref{linear1} to \eqref{e_linear6}, because it just serves as a perturbation to excite the eigenmodes. For simplifications, we make the following assumptions:
\begin{enumerate}
    \item $k_y$ is negligible compared to $k_x$, i.e., the perturbations are nearly axisymmetric, which is suitable for the density waves far away from corotation resonance point and naturally fits the quasi-drift mode.
    \item The gas diffusion terms are ignored. Gas diffusion is not important because of the long-scale nature of gas waves. (It would be worthwhile to stress that the diffusion affect $\Sigma_d/\Sigma_g$ but not $\Sigma_d$)
    \item The gas is not damped by the dust because of the small dust-gas ratio $\mu$, while we reserve its driving terms on dust to relate their perturbations. 
\end{enumerate}
Then the equations read
\begin{equation}
    \begin{bmatrix}
        - i \tilde{\omega}_g  & i k_x & 0 & 0 & 0 & 0 \\
        i k_x & - i \tilde{\omega}_g & - 2 & - \frac{\mu \tilde{w}_{s,x}}{\tau} & - \frac{\mu}{\tau} & 0 \\
        0 & \frac{1}{2} & - i \tilde{\omega}_g &  - \frac{\mu \tilde{w}_{s,y}}{\tau} & 0 & - \frac{\mu}{\tau} \\
        0 & 0 & 0 & - i \tilde{\omega}_d^{\prime } & i k_x & 0 \\
        0 & - \frac{1}{\tau} & 0 & \frac{\alpha i k_x}{\tau} & - i \tilde{\omega}_d^{\prime } + \frac{1}{\tau} & - 2 \\
        0 & 0 & - \frac{1}{\tau} & 0 & \frac{1}{2} & - i \tilde{\omega}_d^{\prime } + \frac{1}{\tau} & 
    \end{bmatrix}
    \begin{bmatrix}
        f_g^{-1} \delta \tilde{\Sigma}_{g}\\
        \delta \tilde{u}_{gx} \\
        \delta \tilde{u}_{gy} \\
        f_d^{-1} \delta \tilde{\Sigma}_{d}  \\
        \delta \tilde{u}_{dx} \\
        \delta \tilde{u}_{dy}
    \end{bmatrix} = 0.
\end{equation}
We represent the equations as $\mathcal{M}{\lambda}=0$, where $\mathcal{M}$ is a $6 \times 6$ matrix and can be represented with four block $3 \times 3$ matrices, i.e.,
\begin{gather}
    \mathcal{M}
    =
    \begin{bmatrix}
        \mathcal{G} & \mathcal{F}_g \\
        \mathcal{F}_d & \mathcal{D}
    \end{bmatrix},
    \mathcal{G}
    =
    \begin{bmatrix}
        - i \tilde{\omega}_g  & i k_x & 0 \\
        i k_x & - i \tilde{\omega}_g & - 2 \\
        0 & \frac{1}{2} & - i \tilde{\omega}_g
    \end{bmatrix},
    \mathcal{F}_d
    =
    \begin{bmatrix}
        0 & 0 & 0 \\
        - \frac{\mu \tilde{w}_{s,x}}{\tau} & - \frac{\mu}{\tau} & 0 \\
        - \frac{\mu \tilde{w}_{s,y}}{\tau} & 0 & - \frac{\mu}{\tau}
    \end{bmatrix}, \\
    \mathcal{F}_g
    =
    \begin{bmatrix}
        0 & 0 & 0 \\
        0 & - \frac{1}{\tau} & 0 \\
        0 & 0 & - \frac{1}{\tau}
    \end{bmatrix},
    \mathcal{D}
    =
    \begin{bmatrix}
        - i \tilde{\omega}_d^{\prime } & i k_x & 0 \\
        \frac{\alpha i k_x}{\tau} & - i \tilde{\omega}_d^{\prime } + \frac{1}{\tau} & - 2 \\
        0 & \frac{1}{2} & - i \tilde{\omega}_d^{\prime } + \frac{1}{\tau} & 
    \end{bmatrix}.
\end{gather}
To get the nontrivial solutions, we need $ \mathrm{det} \left( \mathcal{M} \right) = 0$ with
\begin{equation} \label{Schur}
    \mathrm{det} \left( \mathcal{M} \right) = \mathrm{det} \left( \mathcal{G} \right) \mathrm{det} \left( \mathcal{D} -   \mathcal{F}_g  \mathcal{G}^{-1}  \mathcal{F}_d \right), 
\end{equation}
It is the Schur's formula, where $\mathcal{D} -   \mathcal{F}_g  \mathcal{A}^{-1}  \mathcal{F}_d$ is called the Schur complement of $\mathcal{G}$. We can easily get a physical intuition for the two-fluid system from the formula. Matrices $\mathcal{G}$ and $\mathcal{D}$ delineate the gaseous motion and dusty motion, respectively. Matrices $\mathcal{F}_d$ and $\mathcal{F}_g$ delineate the dusty forcing on gas and gaseous forcing on dust, respectively. When $f_d$ is low, the system includes the modes in the pure gas, i.e., 
\begin{equation}
    \mathrm{det} \left( \mathcal{G} \right) = - i \tilde{\omega}_g \left( 1 - \tilde{\omega}_g^2 + k_x^2 \right) = 0, 
\end{equation}
where $1 - \tilde{\omega}_g^2 + k_x^2=0$ represents gas density waves. The other part of \eqref{Schur} represents the dusty motion modified by gas, i.e., $ \mathrm{det} \left( \mathcal{D} -   \mathcal{F}_g  \mathcal{G}^{-1}  \mathcal{F}_d \right)$. The matrix $ -\mathcal{F}_g  \mathcal{G}^{-1}  \mathcal{F}_d$ just delineates the gaseous driving on dusty motion, where $\mathcal{G}^{-1} = \mathrm{adj} \left( \mathcal{G} \right) \mathrm{det} \left( \mathcal{G} \right)^{-1}$ and $\mathrm{adj}\left(\mathcal{G}\right)$ is the adjoint matrix of $\mathcal{G}$. It is easily to see that the gaseous force is in direct proportion to the excitation of gaseous motion represented by $\mathrm{det} \left( \mathcal{G} \right)^{-1}$. Finally,  $\mathrm{det} \left( \mathcal{D} -   \mathcal{F}_g  \mathcal{G}^{-1}  \mathcal{F}_d \right)=0$  results in
\begin{gather} 
    i \tilde{\omega}_d^{\prime 3} - \tilde{\omega}_d^{\prime 2}\left[ \frac{2}{\tau} + \frac{2\mu \left(\tilde{\omega}_g^{2} - k_c^2 \right)}{|\mathcal{G}|\tau^2} \right] - i \tilde{\omega}_d^{\prime }\left[ \frac{\alpha k_x^2}{\tau} -  \frac{i k_x \mu \tilde{\omega}_g}{|\mathcal{G}|\tau^2} \left( w_{s,x}\tilde{\omega}_g+2i w_{s,y} \right) \right] \nonumber \\
    - i \tilde{\omega}_d^{\prime }\left[ 1 + \frac{1}{\tau^2} + \left( \frac{ \mu \tilde{\omega}_g }{|\mathcal{G}| \tau^2} \right)^2 \left(\tilde{\omega}_g^2 -1\right) + \frac{\mu}{|\mathcal{G}|\tau^2}\left( \frac{2\tilde{\omega}_g^2 - k_c^2}{\tau} - 2i \tilde{\omega}_g \right)\right] \nonumber \\
    +\frac{k_x \mu \tilde{\omega}_g}{|\mathcal{G}| \tau^2}\Bigg\{ -\left\{ 1 + i \tilde{\omega}_g\left[ \frac{1}{\tau} + \frac{\mu \left( \tilde{\omega}_g^2 - k_c^2 -1 \right)}{| \mathcal{G} | \tau^2} \right]-i \right\} w_{s,x} + 2\left( \frac{1}{\tau} -i \tilde{\omega}_g - \frac{\mu k_c^2}{| \mathcal{G} |\tau^2}\right) w_{s,y} \Bigg\} + \frac{\alpha k_x^2}{\tau^2}\left[ 1 + \frac{\mu \left( \tilde{\omega}_g^2 - k_c^2 \right)}{ | \mathcal{G} | \tau} \right] = 0, \label{eq_DR}
\end{gather}
where $k_c$ is the $k_x$ originating from the pressure term in gaseous equations. If we assume $\tilde{\omega}_{d} \simeq \tilde{\omega}_{g}$ is large (when $x$ is large) and $k_x$ becomes small (the terms involving $\alpha$ and $k_x v_{dx}$ are neglected). Then we can write Equation \eqref{eq_DR} as
\begin{equation}
    -i \tilde{\omega}_d  \left[ 1 - \tilde{\omega}_d^2 + f\left( \tau \right)\right] = 0,
\end{equation}
where $f\left(\tau\right)$ represents all the forcing and damping terms from aerodynamic drag. In this case, the dispersion relation closely resembles that of gaseous density waves, with the resilience term replaced by aerodynamic drag.

Now, let us consider $k_x$ and $\omega_d \simeq \omega_g$ large, $\tau \sim 1$, and neglect the terms of lower order, Equation \eqref{eq_DR} reads
\begin{gather}
    \tilde{\omega}_d^{\prime 3} - \left( \alpha k_x^2 + 1 \right) \tilde{\omega}_d^{\prime } - i \alpha k_x^2 = 0,
\end{gather}
where three solutions can be obtained, two of them are important. They are actually split from the quasi-drift mode and have reduced wavenumbers. One of them is ingoing, similar to the quasi-drift mode, contributing a significant positive torque to the planet in most cases. In contrast, the other one is outgoing, resulting in a moderate negative torque on the planet.

Without dust diffusion ($\alpha=0$), Equation \eqref{eq_DR} reads
\begin{gather} 
    i \tilde{\omega}_d^{\prime 3} - \tilde{\omega}_d^{\prime 2}\left[ \frac{2}{\tau} + \frac{2\mu \left(\tilde{\omega}_g^{2} - k_c^2 \right)}{|\mathcal{G}|\tau^2} \right] - \tilde{\omega}_d^{\prime } \frac{ k_x \mu \tilde{\omega}_g}{|\mathcal{G}|\tau^2} \left( w_{s,x}\tilde{\omega}_g+2i w_{s,y} \right) - i \tilde{\omega}_d^{\prime }\left[ 1 + \frac{1}{\tau^2} + \left( \frac{ \mu \tilde{\omega}_g }{|\mathcal{G}| \tau^2} \right)^2 \left(\tilde{\omega}_g^2 -1\right) + \frac{\mu}{|\mathcal{G}|\tau^2}\left( \frac{2\tilde{\omega}_g^2 - k_c^2}{\tau} - 2i \tilde{\omega}_g \right)\right] \nonumber \\
    +\frac{k_x \mu \tilde{\omega}_g}{|\mathcal{G}| \tau^2}\Bigg\{ -\left\{ 1 + i \tilde{\omega}_g\left[ \frac{1}{\tau} + \frac{\mu \left( \tilde{\omega}_g^2 - k_c^2 -1 \right)}{| \mathcal{G} | \tau^2} \right]-i \right\} w_{s,x} + 2\left( \frac{1}{\tau} -i \tilde{\omega}_g - \frac{\mu k_c^2}{| \mathcal{G} |\tau^2}\right) w_{s,y} \Bigg\} = 0, \label{eq_noalpha}
\end{gather}
where the last term, proportional to $|\mathcal{G}|^{-1}$, originates from the gas drag. Without that term, the free drift mode can be easily recovered by the common term $\tilde{\omega}_d^{\prime}= \tilde{\omega}_d - k_x v_{dx} = 0 $, while the gas drag modifies it to a quasi-drift mode.

\bibliography{ST_diff.bib}{}

\begin{thebibliography}{}
\expandafter\ifx\csname natexlab\endcsname\relax\def\natexlab#1{#1}\fi
\providecommand{\url}[1]{\href{#1}{#1}}
\providecommand{\dodoi}[1]{doi:~\href{http://doi.org/#1}{\nolinkurl{#1}}}
\providecommand{\doeprint}[1]{\href{http://ascl.net/#1}{\nolinkurl{http://ascl.net/#1}}}
\providecommand{\doarXiv}[1]{\href{https://arxiv.org/abs/#1}{\nolinkurl{https://arxiv.org/abs/#1}}}

\bibitem[{{Armitage}(2020)}]{Armitage2020}
{Armitage}, P.~J. 2020, {Astrophysics of planet formation, Second Edition}

\bibitem[{{Baruteau} \& {Lin}(2010)}]{Baruteau2010}
{Baruteau}, C., \& {Lin}, D.~N.~C. 2010, \apj, 709, 759,
  \dodoi{10.1088/0004-637X/709/2/759}

\bibitem[{{Ben{\'\i}tez-Llambay} \& {Pessah}(2018)}]{Llambay2018}
{Ben{\'\i}tez-Llambay}, P., \& {Pessah}, M.~E. 2018, \apjl, 855, L28,
  \dodoi{10.3847/2041-8213/aab2ae}

\bibitem[{{Brown} \& {Ogilvie}(2024)}]{Brown2024}
{Brown}, J.~J., \& {Ogilvie}, G.~I. 2024, \mnras, 534, 39,
  \dodoi{10.1093/mnras/stae2060}

\bibitem[{{Chen} \& {Lin}(2020)}]{Chen2020}
{Chen}, K., \& {Lin}, M.-K. 2020, \apj, 891, 132,
  \dodoi{10.3847/1538-4357/ab76ca}

\bibitem[{{Chrenko} {et~al.}(2024){Chrenko}, {Chametla}, {Masset}, {Baruteau},
  \& {Bro{\v{z}}}}]{Chrenko2024}
{Chrenko}, O., {Chametla}, R.~O., {Masset}, F.~S., {Baruteau}, C., \&
  {Bro{\v{z}}}, M. 2024, \aap, 690, A41, \dodoi{10.1051/0004-6361/202450922}

\bibitem[{{D'Angelo} \& {Lubow}(2010)}]{D'Angelo2010}
{D'Angelo}, G., \& {Lubow}, S.~H. 2010, \apj, 724, 730,
  \dodoi{10.1088/0004-637X/724/1/730}

\bibitem[{{Dubrulle} {et~al.}(1995){Dubrulle}, {Morfill}, \&
  {Sterzik}}]{Dubrulle1995}
{Dubrulle}, B., {Morfill}, G., \& {Sterzik}, M. 1995, \icarus, 114, 237,
  \dodoi{10.1006/icar.1995.1058}

\bibitem[{{Dullemond} \& {Penzlin}(2018)}]{Dullemond2018}
{Dullemond}, C.~P., \& {Penzlin}, A.~B.~T. 2018, \aap, 609, A50,
  \dodoi{10.1051/0004-6361/201731878}

\bibitem[{{Fairbairn} \& {Rafikov}(2024)}]{Fairbairn2024}
{Fairbairn}, C.~W., \& {Rafikov}, R.~R. 2024, arXiv e-prints, arXiv:2407.20398,
  \dodoi{10.48550/arXiv.2407.20398}

\bibitem[{{Goldreich} \& {Lynden-Bell}(1965)}]{Goldreich1965}
{Goldreich}, P., \& {Lynden-Bell}, D. 1965, \mnras, 130, 125,
  \dodoi{10.1093/mnras/130.2.125}

\bibitem[{{Goldreich} \& {Tremaine}(1979)}]{Goldreich1979}
{Goldreich}, P., \& {Tremaine}, S. 1979, \apj, 233, 857, \dodoi{10.1086/157448}

\bibitem[{{Gole} {et~al.}(2020){Gole}, {Simon}, {Li}, {Youdin}, \&
  {Armitage}}]{Gole2020}
{Gole}, D.~A., {Simon}, J.~B., {Li}, R., {Youdin}, A.~N., \& {Armitage}, P.~J.
  2020, \apj, 904, 132, \dodoi{10.3847/1538-4357/abc334}

\bibitem[{Goodman \& Rafikov(2001)}]{Goodman2001}
Goodman, J., \& Rafikov, R.~R. 2001, The Astrophysical Journal, 552, 793,
  \dodoi{10.1086/320572}

\bibitem[{{Harris} {et~al.}(2020){Harris}, {Millman}, {van der Walt},
  {Gommers}, {Virtanen}, {Cournapeau}, {Wieser}, {Taylor}, {Berg}, {Smith},
  {Kern}, {Picus}, {Hoyer}, {van Kerkwijk}, {Brett}, {Haldane}, {del R{\'\i}o},
  {Wiebe}, {Peterson}, {G{\'e}rard-Marchant}, {Sheppard}, {Reddy}, {Weckesser},
  {Abbasi}, {Gohlke}, \& {Oliphant}}]{Harris2020}
{Harris}, C.~R., {Millman}, K.~J., {van der Walt}, S.~J., {et~al.} 2020, \nat,
  585, 357, \dodoi{10.1038/s41586-020-2649-2}

\bibitem[{{Hayashi}(1981)}]{Hayashi1981}
{Hayashi}, C. 1981, Progress of Theoretical Physics Supplement, 70, 35,
  \dodoi{10.1143/PTPS.70.35}

\bibitem[{{Hopkins} \& {Squire}(2018)}]{Hopkins2018}
{Hopkins}, P.~F., \& {Squire}, J. 2018, \mnras, 480, 2813,
  \dodoi{10.1093/mnras/sty1982}

\bibitem[{{Hou} \& {Yu}(2024)}]{Hou2024}
{Hou}, Q., \& {Yu}, C. 2024, \apj, 972, 152, \dodoi{10.3847/1538-4357/ad6a5c}

\bibitem[{{Hsieh} \& {Lin}(2020)}]{Hsieh2020}
{Hsieh}, H.-F., \& {Lin}, M.-K. 2020, \mnras, 497, 2425,
  \dodoi{10.1093/mnras/staa2115}

\bibitem[{{Huang} \& {Bai}(2022)}]{HPH2022}
{Huang}, P., \& {Bai}, X.-N. 2022, \apjs, 262, 11,
  \dodoi{10.3847/1538-4365/ac76cb}

\bibitem[{{Huang} \& {Yu}(2022)}]{Huang2022}
{Huang}, S., \& {Yu}, C. 2022, \mnras, 514, 1733,
  \dodoi{10.1093/mnras/stac1464}

\bibitem[{{Hunter}(2007)}]{Hunter2007}
{Hunter}, J.~D. 2007, Computing in Science and Engineering, 9, 90,
  \dodoi{10.1109/MCSE.2007.55}

\bibitem[{{Klahr} \& {Schreiber}(2021)}]{Klahr2021}
{Klahr}, H., \& {Schreiber}, A. 2021, \apj, 911, 9,
  \dodoi{10.3847/1538-4357/abca9b}

\bibitem[{{Korycansky} \& {Pollack}(1993)}]{Korycansky1993}
{Korycansky}, D.~G., \& {Pollack}, J.~B. 1993, \icarus, 102, 150,
  \dodoi{10.1006/icar.1993.1039}

\bibitem[{{Lambrechts} \& {Johansen}(2012)}]{Lambrechts2012}
{Lambrechts}, M., \& {Johansen}, A. 2012, \aap, 544, A32,
  \dodoi{10.1051/0004-6361/201219127}

\bibitem[{{Latter} \& {Rosca}(2017)}]{Latter2017}
{Latter}, H.~N., \& {Rosca}, R. 2017, \mnras, 464, 1923,
  \dodoi{10.1093/mnras/stw2455}

\bibitem[{{Lee}(2024)}]{Lee2024}
{Lee}, E.~J. 2024, \apjl, 970, L15, \dodoi{10.3847/2041-8213/ad5d8e}

\bibitem[{{Liu} \& {Bai}(2023)}]{Liu2023}
{Liu}, H., \& {Bai}, X.-N. 2023, \mnras, 526, 80,
  \dodoi{10.1093/mnras/stad2629}

\bibitem[{{Magnan} {et~al.}(2024{\natexlab{a}}){Magnan}, {Heinemann}, \&
  {Latter}}]{Magnan2024a}
{Magnan}, N., {Heinemann}, T., \& {Latter}, H.~N. 2024{\natexlab{a}}, \mnras,
  \dodoi{10.1093/mnras/stae1978}

\bibitem[{{Magnan} {et~al.}(2024{\natexlab{b}}){Magnan}, {Heinemann}, \&
  {Latter}}]{Magnan2024b}
---. 2024{\natexlab{b}}, \mnras, 529, 688, \dodoi{10.1093/mnras/stae052}

\bibitem[{{Masset} \& {Papaloizou}(2003)}]{Masset2003}
{Masset}, F.~S., \& {Papaloizou}, J.~C.~B. 2003, \apj, 588, 494,
  \dodoi{10.1086/373892}

\bibitem[{{Nakagawa} {et~al.}(1986){Nakagawa}, {Sekiya}, \&
  {Hayashi}}]{Nakagawa1986}
{Nakagawa}, Y., {Sekiya}, M., \& {Hayashi}, C. 1986, \icarus, 67, 375,
  \dodoi{10.1016/0019-1035(86)90121-1}

\bibitem[{{Narayan} {et~al.}(1987){Narayan}, {Goldreich}, \&
  {Goodman}}]{Narayan1987}
{Narayan}, R., {Goldreich}, P., \& {Goodman}, J. 1987, \mnras, 228, 1,
  \dodoi{10.1093/mnras/228.1.1}

\bibitem[{{Ogilvie} \& {Lubow}(2006)}]{Ogilvie2006}
{Ogilvie}, G.~I., \& {Lubow}, S.~H. 2006, \mnras, 370, 784,
  \dodoi{10.1111/j.1365-2966.2006.10506.x}

\bibitem[{{Ormel} \& {Klahr}(2010)}]{Ormel2010}
{Ormel}, C.~W., \& {Klahr}, H.~H. 2010, \aap, 520, A43,
  \dodoi{10.1051/0004-6361/201014903}

\bibitem[{{Paardekooper}(2014)}]{Paardekooper2014}
{Paardekooper}, S.~J. 2014, \mnras, 444, 2031, \dodoi{10.1093/mnras/stu1542}

\bibitem[{{Paardekooper} {et~al.}(2020){Paardekooper}, {McNally}, \&
  {Lovascio}}]{Paardekooper2020}
{Paardekooper}, S.-J., {McNally}, C.~P., \& {Lovascio}, F. 2020, \mnras, 499,
  4223, \dodoi{10.1093/mnras/staa3162}

\bibitem[{{Press} {et~al.}(1992){Press}, {Teukolsky}, {Vetterling}, \&
  {Flannery}}]{Press1992}
{Press}, W.~H., {Teukolsky}, S.~A., {Vetterling}, W.~T., \& {Flannery}, B.~P.
  1992, {Numerical recipes in FORTRAN. The art of scientific computing}

\bibitem[{{Rafikov} \& {Petrovich}(2012)}]{Rafikov2012}
{Rafikov}, R.~R., \& {Petrovich}, C. 2012, \apj, 747, 24,
  \dodoi{10.1088/0004-637X/747/1/24}

\bibitem[{{Reg{\'a}ly}(2020)}]{Regaly2020}
{Reg{\'a}ly}, Z. 2020, \mnras, 497, 5540, \dodoi{10.1093/mnras/staa2181}

\bibitem[{{Shadmehri}(2016)}]{Shadmehri2016}
{Shadmehri}, M. 2016, \apj, 817, 140, \dodoi{10.3847/0004-637X/817/2/140}

\bibitem[{{Shakura} \& {Sunyaev}(1973)}]{Shakura1973}
{Shakura}, N.~I., \& {Sunyaev}, R.~A. 1973, \aap, 24, 337

\bibitem[{{Squire} \& {Hopkins}(2018{\natexlab{a}})}]{Squire2018a}
{Squire}, J., \& {Hopkins}, P.~F. 2018{\natexlab{a}}, \apjl, 856, L15,
  \dodoi{10.3847/2041-8213/aab54d}

\bibitem[{{Squire} \& {Hopkins}(2018{\natexlab{b}})}]{Squire2018b}
---. 2018{\natexlab{b}}, \mnras, 477, 5011, \dodoi{10.1093/mnras/sty854}

\bibitem[{{Takahashi} \& {Inutsuka}(2014)}]{Takahashi2014}
{Takahashi}, S.~Z., \& {Inutsuka}, S.-i. 2014, \apj, 794, 55,
  \dodoi{10.1088/0004-637X/794/1/55}

\bibitem[{{Takeuchi} \& {Miyama}(1998)}]{Takeuchi1998}
{Takeuchi}, T., \& {Miyama}, S.~M. 1998, \pasj, 50, 141,
  \dodoi{10.1093/pasj/50.1.141}

\bibitem[{{Tanaka} \& {Okada}(2024)}]{Tanaka2024}
{Tanaka}, H., \& {Okada}, K. 2024, arXiv e-prints, arXiv:2404.12521,
  \dodoi{10.48550/arXiv.2404.12521}

\bibitem[{{Tanaka} {et~al.}(2002){Tanaka}, {Takeuchi}, \& {Ward}}]{Tanaka2002}
{Tanaka}, H., {Takeuchi}, T., \& {Ward}, W.~R. 2002, \apj, 565, 1257,
  \dodoi{10.1086/324713}

\bibitem[{{Tominaga} {et~al.}(2021){Tominaga}, {Inutsuka}, \&
  {Kobayashi}}]{Tominaga2021}
{Tominaga}, R.~T., {Inutsuka}, S.-i., \& {Kobayashi}, H. 2021, \apj, 923, 34,
  \dodoi{10.3847/1538-4357/ac173a}

\bibitem[{{Tominaga} {et~al.}(2023){Tominaga}, {Inutsuka}, \&
  {Takahashi}}]{Tominaga2023}
{Tominaga}, R.~T., {Inutsuka}, S.-i., \& {Takahashi}, S.~Z. 2023, \apj, 953,
  60, \dodoi{10.3847/1538-4357/ace043}

\bibitem[{{Tominaga} {et~al.}(2022{\natexlab{a}}){Tominaga}, {Kobayashi}, \&
  {Inutsuka}}]{Tominaga2022a}
{Tominaga}, R.~T., {Kobayashi}, H., \& {Inutsuka}, S.-i. 2022{\natexlab{a}},
  \apj, 937, 21, \dodoi{10.3847/1538-4357/ac82b4}

\bibitem[{{Tominaga} {et~al.}(2019){Tominaga}, {Takahashi}, \&
  {Inutsuka}}]{Tominaga2019}
{Tominaga}, R.~T., {Takahashi}, S.~Z., \& {Inutsuka}, S.-i. 2019, \apj, 881,
  53, \dodoi{10.3847/1538-4357/ab25ea}

\bibitem[{{Tominaga} {et~al.}(2022{\natexlab{b}}){Tominaga}, {Tanaka},
  {Kobayashi}, \& {Inutsuka}}]{Tominaga2022b}
{Tominaga}, R.~T., {Tanaka}, H., {Kobayashi}, H., \& {Inutsuka}, S.-i.
  2022{\natexlab{b}}, \apj, 940, 152, \dodoi{10.3847/1538-4357/ac97e8}

\bibitem[{{Umurhan} {et~al.}(2020){Umurhan}, {Estrada}, \&
  {Cuzzi}}]{Umurhan2020}
{Umurhan}, O.~M., {Estrada}, P.~R., \& {Cuzzi}, J.~N. 2020, \apj, 895, 4,
  \dodoi{10.3847/1538-4357/ab899d}

\bibitem[{{Ward}(1997)}]{Ward1997}
{Ward}, W.~R. 1997, \icarus, 126, 261, \dodoi{10.1006/icar.1996.5647}

\bibitem[{{Weidenschilling}(1977)}]{Weidenschilling1977}
{Weidenschilling}, S.~J. 1977, \mnras, 180, 57, \dodoi{10.1093/mnras/180.2.57}

\bibitem[{{Wu} \& {Chen}(2025)}]{Wu2025}
{Wu}, Y., \& {Chen}, Y.-X. 2025, \mnras, 536, L13,
  \dodoi{10.1093/mnrasl/slae102}

\bibitem[{{Youdin} \& {Goodman}(2005)}]{Youdin2005}
{Youdin}, A.~N., \& {Goodman}, J. 2005, \apj, 620, 459, \dodoi{10.1086/426895}

\bibitem[{{Yu} {et~al.}(2010){Yu}, {Li}, {Li}, {Lubow}, \& {Lin}}]{Yu2010}
{Yu}, C., {Li}, H., {Li}, S., {Lubow}, S.~H., \& {Lin}, D.~N.~C. 2010, \apj,
  712, 198, \dodoi{10.1088/0004-637X/712/1/198}

\end{thebibliography}
\bibliographystyle{aasjournal}
\end{document}